\documentclass[twoside, 11pt]{article}

\usepackage[usenames,dvipsnames,table]{xcolor}

\usepackage{blindtext}
\usepackage[abbrvbib, preprint]{jmlr2e}
\usepackage{lastpage}

\usepackage{float,csquotes,xcolor}
\usepackage{stfloats}
\usepackage{graphicx}
\usepackage{amsfonts}
\usepackage{natbib}
\usepackage{amsmath}
\usepackage{bm}
\usepackage{bbm}

\usepackage{tikz}

\usepackage[linesnumbered]{algorithm2e}
\usepackage{algpseudocode}
\RestyleAlgo{ruled}

\usepackage{mathtools}
\DeclarePairedDelimiter{\ceil}{\lceil}{\rceil}

\newcommand{\R}{\mathbb{R}}

\hypersetup{
    colorlinks=true,
    linkcolor=red,
    citecolor=MidnightBlue,
    }

\definecolor{Gray}{gray}{0.9}

\jmlrheading{23}{2022}{1-\pageref{LastPage}}{TBD; Revised TBD}{TBD}{21-0000}{Margossian, Zhong and Mukherjee}

\ShortHeadings{Potts models simulation}{Margossian, Zhong and Mukherjee}
\firstpageno{1}

\definecolor{Gray}{gray}{0.9}

\begin{document}


\title{Efficient Sampling for Ising and Potts Models \\ using Auxiliary Gaussian Variables}

\author{\name Charles C. Margossian\textcolor{red}{$^*$} \email charles.margossian@ubc.ca \\
  \addr Department of Statistics \\
  University of British Columbia \\
  Vancouver, BC, Canada
  \AND
  \name Chenyang Zhong\textcolor{red}{$^*$} \email cz2755@columbia.edu \\
  \addr Department of Statistics \\
  Columbia University \\
  New York, NY, USA
  \AND
  \name Sumit Mukherjee \email sm3949@columbia.edu \\
  \addr Department of Statistics \\
  Columbia University \\
  New York, NY, USA \\[0.1in]
  \textcolor{red}{*}Equal contribution
  }

\editor{TBD}

\maketitle

\begin{abstract}
Ising and Potts models are an important class of discrete probability distributions which originated from statistical physics and since then have found applications in several disciplines. 
Simulation from these models is a well known challenging problem. 
In this paper, we study a class of Markov chain Monte Carlo algorithms, in which we introduce an auxiliary Gaussian variable such that, conditional on this variable, the discrete states are independent.
This approach is broadly applicable to Ising and Potts models, including ones in which the coupling matrix admits negative entries, as in spin glass and Hopfield models.
We focus on a block Gibbs sampler version of this algorithm, which alternates between sampling the auxiliary Gaussian and the discrete states, and derive mixing time bounds for a wide class of Ising/Potts models at both high and low temperatures, yielding results analogous to those derived for the Heat Bath and Swendsen-Wang algorithms. We present novel choices of auxiliary Gaussian variables which scale well with the number of states in the Potts model, and which can take advantage of the low rank structure of the coupling matrix, if any. 
Finally, we numerically evaluate the performance of the auxiliary Gaussian Gibbs sampler with several competing algorithms,  across a range of examples.
\end{abstract}

\ \\[-0.5pt]
\begin{keywords}
Ising models, Potts models, Markov chain Monte Carlo, Gibbs sampling, mixing time
\end{keywords}


\section{Introduction} \label{sec:intro}

The Ising model is a probability distribution on the space of binary vectors of size $n$, whose components are allowed to take two values, traditionally taken to be $\{0, 1\}$ or $\{-1, 1\}$.
The Ising model was first introduced in statistical physics to study ferromagnetism \citep{Ising:1925}.
The Potts model \citep{Potts:1952} generalizes the Ising model to $q \ge 2$ states, where the states can be taken to be $[q] := \{1, 2, \cdots, q\}$.
With this choice, the Potts model is given by the following probability mass function on $[q]^n$:
\begin{equation} \label{eq:potts}
  \mathbb P(\mathbf X = \mathbf x) =\frac{1}{Z(\beta,A) }\exp \left (\frac{\beta}{2}\sum_{i, j = 1}^n A_{ij} 1 \{x_i = x_j \} \right ).
\end{equation}
Here $\beta>0$ is the inverse temperature parameter; the coupling matrix $A$ is a symmetric matrix which controls the dependence between the components of ${\bf X}$,  and $Z(\beta,A)$ is the normalizing constant,
also termed the partition function.
We assume that the diagonal entries of $A$ equal $0$, noting that this has no impact on the model \eqref{eq:potts}.
The most common choice of $A$ is as a scaled adjacency matrix of a graph on $n$ vertices. 
We consider both positive and negative entries for $A$, thus covering both traditional Ising/Potts models and spin glass models under one framework. 

Since their inception, the Ising and Potts models have found applications in a wide range of disciplines, including image processing, protein folding, neuroscience, and in the social sciences; see for example \citet{Lipowski:2022} and references therein.
Two prominent cases of Ising models in machine learning are the Hopfield network and the Boltzmann machine \citep{Mackay:2003}.
One of the main reasons for the popularity of Ising and Potts models is that they are perhaps the simplest models which exhibit non-trivial dependence across their components. 

Simulation plays a crucial role in understanding the properties of the Ising and Potts models, particularly through the computation of Monte Carlo estimators.
Building efficient samplers for such models is a well-known challenge \citep{Neal:1993, Landau:2009, Faulkner:2024}.
Over the past decade, this problem has generated much interest in the machine learning community \citep[e.g][]{Martens:2010, Zhang:2012, Pakman:2013, Grathwohl:2021, Robnik:2023}.
A general strategy, first proposed by \citet{Martens:2010}, is to introduce an auxiliary Gaussian (AG) variable ${\bf Z}$, such that the elements of ${\bf X}$ conditional on ${\bf Z}$ are mutually independent.
It is then possible to construct Markov chain Monte Carlo (MCMC) algorithms, which sample from the joint space $({\bf X}, {\bf Z})$.
For example, we may implement a block Gibbs sampler that alternates between sampling from ${\bf X} \mid {\bf Z}$ and ${\bf Z} \mid {\bf X}$.
We refer to such an algorithm as an AG sampler.
The idea of introducing a latent continuous variable to study discrete states in an Ising model can be further traced back to the Hubbard–Stratonovich transformation used in theoretical physics \citep{Stratonovich:1957, Hubbard:1959}.

The first contribution of our paper is algorithmic.
We introduce two novel constructions of AG variables that in turn lead to new sampling algorithms.
Specifically, we present a construction for Potts models and another construction for (approximately) low-rank coupling matrices $A$, with applications to dense regular graphs and Hopfield networks.
These constructions can be used separately or together.

The second contribution of our paper is theoretical and concerns bounds on the mixing time of the AG sampler.
To the best of our knowledge, these are the first mixing time bounds for AG samplers.
Such bounds exist for classical samplers, including the heat bath \citep[e.g.][]{Levin:2010, Cuff:2012} and the Swendsen-Wang algorithm \citep[e.g.][]{Long:2014, Gheissari:2018, Galanis:2019, Blanca:2021}.
The study of these bounds mainly focuses on cases where $\beta$ is small (high-temperature regime) and/or $A$ is the adjacency matrix of a complete graph (Curie-Weiss model).
Our bounds demonstrate fast mixing of the AG sampler at high temperatures for a wide class of coupling matrices, including Ising and Potts models on graphs and spin glass models such as the Hopfield model.
In some limited cases, such as the Curie-Weiss model, we also derive mixing bounds at low temperatures.

Finally, our paper contains extensive numerical experiments, which demonstrate the performance of AG samplers relative to existing algorithms.
We examine a number of target distributions, including Potts models on graphs, Hopfield networks and spin glass models.
We consider a range of temperatures, including settings where the system is near a critical temperature or cold.
In many regimes, we find that the AG sampler yields the best performance.

\section{Choices of Auxiliary Gaussian} \label{sec:choice}

In this section, we derive the  auxiliary Gaussian algorithm for the Potts model.
We then discuss a modification of the algorithm for low-rank coupling matrices $A$ that achieves higher computational efficiency.

\subsection{Auxiliary Gaussian for Potts} \label{sec:PottsAG}

In order to derive an auxiliary Gaussian variable for the Potts model, it is useful to rewrite the probability mass function in terms of one-hot variables.
Consider a vector ${\mathbf x}=(x_1,\ldots,x_n)\in [q]^n$.
For every $\ell\in [q]$, define the one-hot vector,
$${\mathbf y}_\ell:=(1\{x_1=\ell\}, \ldots, 1\{x_n=\ell\})'\in \R^n.$$
In other words, the $i^\text{th}$ element of $\mathbf y_\ell$ is 1 if $x_i=\ell$, and $0$ otherwise.
This sets up a one-to-one map between $\mathbf x$ and $({\bf y}_1,\ldots,{\bf y}_q)$.
The log probability mass function of the Potts model may then be written as
\begin{equation*}
\frac{\beta}{2}  \sum_{i, j = 1}^n A_{ij} 1 \{x_i = x_j \} =\frac{\beta}{2}  \sum_{\ell = 1}^q \mathbf y'_\ell A \mathbf y_\ell,
\end{equation*}
where the additive log normalizing constant has been dropped.
The term $\mathbf y'_\ell A \mathbf y_\ell / 2$ looks like the log moment generating function of a multivariate Gaussian, except that the matrix $A$ is not non-negative definite, as the diagonal entries of $A$ equal $0$.
Let $\lambda_\text{min}$ be the smallest eigenvalue of $A$ and consider $\lambda$ such that $\lambda > -\lambda_\text{min}$.
We replace $A$ by $A+\lambda I$, where $I \in \mathbb R^{n \times n}$ is the identity matrix, thus ensuring that $A+\lambda I$ is positive definite.
This change to the coupling matrix does not change the probability mass function.
Indeed,
%
%
\begin{equation} \label{eq:change-A}
\sum_{\ell = 1}^q \mathbf y_\ell' (A+\lambda I)\mathbf y_\ell -  \sum_{\ell = 1}^q \mathbf y_\ell' A\mathbf y_\ell
   = \lambda \sum_{\ell= 1}^q \mathbf y_\ell' \mathbf y_\ell
  =\lambda \sum_{\ell=1}^q\sum_{i=1}^n 1\{x_i=\ell\}=n\lambda,
\end{equation}
where the last equality follows from the fact $\sum_{\ell=1}^q 1\{x_i=\ell\}=1$ for each $i\in [n]$,
and the constant $n \lambda$ can be absorbed into the normalizing constant.

Setting 
\begin{equation}\label{DefB}
    B :=  \beta (A + \lambda I),
\end{equation}
the marginal probability mass function of $\mathbf X$ can be written as
\begin{equation}\label{eq:potts_simplified}
\mathbb P(\mathbf X=\mathbf x)=\frac{1}{Z(\beta,B)} \exp\Big(\frac{1}{2}\sum_{\ell=1}^q \mathbf y_\ell'B\mathbf y_\ell\Big).
\end{equation}
We now introduce our auxiliary Gaussian variables.
Given ${\mathbf X}={\mathbf x}$, let $\mathbf Z_1, \ldots, \mathbf Z_q$ be mutually independent $n$-dimensional Gaussian random vectors with
\begin{align}\label{eq:main}
 \mathbf Z_\ell\sim N\left(\mathbf y_\ell, B^{-1}\right).
\end{align}
Then one can check that given ${\mathbf Z}:=({\mathbf Z}_1,\ldots,{\mathbf Z}_q)$  the random variables $(X_1,\ldots,X_n)$ are mutually independent, with
\begin{equation}\label{eq:conditional}
  \mathbb P(X_i = \ell  \mid {\mathbf Z}) =\frac{ \exp \left ( \sum_{j = 1}^n B_{ij} z_{\ell j} \right )}{\sum_{\ell'=1}^q\exp \left ( \sum_{j = 1}^n B_{ij} z_{\ell'j} \right )}.
\end{equation}
By iterative sampling from \eqref{eq:main} and \eqref{eq:conditional}, we obtain a block Gibbs sampler.
This approach extends the block Gibbs sampler proposed by \citet{Martens:2010} for Ising models.
We might also consider running MCMC directly over ${\bf Z}$ and then sampling ${\bf X} \mid {\bf Z}$.
This is the strategy proposed by \citet{Zhang:2012} for the Ising models.
To do so, we need the unnormalized marginal density of ${\bf Z}$, which is given by
\begin{align}\label{eq:marginal}
 p({{\bf z}_1, \cdots, {\bf z}_q}) \propto \exp\Big(-\frac{1}{2}\sum_{\ell=1}^q {\mathbf z}_\ell' B{\mathbf z}_\ell\Big)\prod_{i=1}^n \Big(\sum_{\ell=1}^q \exp\Big(\sum_{j=1}^n B_{ij}z_{\ell  j}\Big) \Big).
\end{align}
The proofs of both \eqref{eq:conditional} and \eqref{eq:marginal} are in Appendix \ref{sec:choice:1}.

\subsection{Block Gibbs using an auxiliary Gaussian}

In this section, we analyze the complexity of the block Gibbs sampler, obtained by alternating between \eqref{eq:main} and \eqref{eq:conditional}.
We refer to this algorithm as the \textit{AG sampler}.
Algorithm~\ref{algo:gibbs} provides details for the AG sampler.
There and throughout the paper, we denote the $t^{th}$ iteration of the Markov chain as ${\bf x}(t)$.
If we look at the $j^\text{th}$ component, we write $x_j(t)$.

The first conditional \eqref{eq:main} requires sampling from a multivariate Gaussian distribution for each $\ell \in [q]$.
Each ${\bf Z}_\ell$ follows a Gaussian with the same covariance matrix $B$ and so the cost of sampling is dominated by the Cholesky decomposition and inversion of $B$, an operation with complexity $\mathcal O(n^3)$.
However, $B$ remains unchanged between iterations, so the decomposition needs be done only once.
Then, within each iteration, the cost is dominated by $2q$ matrix-vector multiplications, each with cost $\mathcal O(n^2)$ (Algorithm~\ref{algo:gibbs} lines 9 and 11). 
Suppose we run MCMC for $T$ iterations.
Then the complexity of the algorithm is $\mathcal O(n^3 + Tn^2q)$.
Provided $n \ll T$, the cost of the initial Cholesky decomposition is marginal. 

\begin{algorithm}
\caption{Block Gibbs with an auxiliary Gaussian (AG sampler)} \label{algo:gibbs}
{\bf Input:} coupling matrix $A$, inverse temperature $\beta$, pertubation $\epsilon > 0$, initialization $\mathbf x(0)$, number of iterations $T$ \\
$\lambda \gets |\lambda_\text{min}(A)| + \epsilon,  \ \epsilon > 0$ \Comment $\lambda_\text{min}(A)$ is the smallest eigenvalue of $A$. \\
$B \gets \beta (A + \lambda\mathbf I)$ \\
$L \gets \text{Cholesky-decompose}(B^{-1})$ \\
\For {$t$ in 1:$T$}{
  \For {$\ell$ in 1:$q$}{
  $\mathbf y_\ell \gets \left (1 \left \{x_1(t - 1) = \ell \right \}, ..., 1 \left \{x_n(t - 1) = \ell \right \} \right)$ \\
  $\mathbf z^*_\ell \sim \text{Normal}(0, I)$ \\
  $\mathbf z_\ell = L_n \mathbf z^*_\ell + \mathbf y_\ell$ \\
  }
  $ P \gets B [\mathbf z_1, ..., \mathbf z_q]$ \\
  \For{j in 1:n}{
  $\textbf p_j \gets \exp(P_{j 1:q})$ \\
  $x_j(t) \sim \text{Categorical}(\textbf p_j / \sum_{\ell = 1}^q p_{j \ell})$
  }
  Draw permutation $\tau_t$ over $[q]$. \Comment Optional state permutation (Section~\ref{sec:permutation}). \\
  ${\bf x}(t) \gets \tau_t({\bf x}(t))$ 
  
}
\textbf{return:} $\mathbf x(1), \mathbf x(2), ..., \mathbf x(T)$
\end{algorithm}

\begin{remark}
  \citet{Martens:2010} propose an auxiliary Gaussian variable over $\mathbb R^{qn}$ for simulating Potts states.
  Plugging this choice into the AG sampler results in an algorithm with complexity $\mathcal O(n^3 + Tn^2q^2)$.
  By contrast, we propose using $q$ auxiliary Gaussian variables, each over $\mathbb R^n$, and the cost per iteration of our sampler scales linearly in $q$, rather than quadratically.
\end{remark}

\subsection{State permutation step} \label{sec:permutation}

The Potts model defined in \eqref{eq:potts} does not prefer any state $\ell \in [q]$ over another, and a permutation of the states does not change the probability mass.
However, at cold temperatures, a Markov chain may take a long time to move from one configuration where a state $\ell$ dominates to one where a state $\ell'$ dominates.
To remedy this issue, we may introduce a random permutation at each iteration $\tau_t$ and do an update
\begin{equation} \label{eq:permuation}
    x_{i}(t ) \leftarrow \tau_t(x_i(t)).
\end{equation}
This update improves Monte Carlo estimates of quantities which vary under state permutations, for example the expected number of particles in a particular state $\ell$, $ \mathbb E \left (\sum_{i} 1\{x_i = \ell \} \right)$.
However, it does not improve estimates of quantities invariant to these permutations, such as the system's expected Hamiltonian, $\mathbb E \left (\beta \sum_{i,j} A_{ij} 1\{x_i = x_j \} /2 \right)$.

The permutation step is also a theoretical tool for bounding mixing times at cold temperatures because the bounded total variation distance accounts for functions which are sensitive to state permutations.
In particular, the permutation step plays a role in Theorem~\ref{Thm3.3}, but it is not used to obtain bounds at high temperatures.

\begin{remark}
    It is possible to break the symmetry over states $\ell \in [q]$ by introducing a magnetic field in \eqref{eq:potts}.
    Then \eqref{eq:permuation} must be treated as a Metropolis proposal, accepted with probability $\text{min} [1, \mathbb P({\bf X} = \tau(x(t))) / \mathbb P({\bf X} = x(t))]$, in order to preserve the correctness of the sampler.
    In the absence of a magnetic component, the acceptance probability is always 1.
\end{remark}

\subsection{Low-rank auxiliary Gaussian} \label{sec:lowrankAG}

For certain models, the modified coupling matrix $B$, obtained after increasing the diagonal elements of $A$, may be low-rank, either exactly or approximately. 
In this case, it is reasonable to believe that we don't have to work with $n$-dimensional Gaussian vectors but instead with a $k$-dimensional Gaussian vector, where $k= {\rm rank}(B)$.

Such algorithms have already been explored in the setting of Ising models for the complete graph \citep{Mukherjee:2018}, where 
\begin{equation}\label{coupling}
A= \frac{1}{n}{\bf 1}{\bf 1}' - \frac{1}{n} I,
\end{equation}
has eigenvalues $(1-1/n,-1/n,\ldots,-1/n)$, where the multiplicity of the eigenvalue $-1/n$ is $n-1$. 
Even though ${\rm rank}(A)=n$ in a strict sense, we can add $I/n$ to $A$ to get the matrix ${\bf 1}{\bf 1}'/n$, which has eigenvalues $(1,0,\ldots,0)$, and rank 1. 
A similar approach was successful in analyzing the Ising model of a regular graph with degree $\propto \sqrt{n}$ \citep{Mukherjee:2021}, where ${\rm rank}(A)\propto \sqrt{n}$.

To propose a general version of the low-rank algorithm, let $B=\beta(A + |\lambda_{\min}| I)$. 
Note it is sufficient to augment the diagonal elements with $|\lambda_{\min}|$, rather than $\lambda > |\lambda_{\min}|$.
Let $\sum_{i=1}^n \mu_i{\bf p}_i{\bf p}_i'$ denote the spectral expansion of $B$, where $(\mu_1,\ldots,\mu_n)$ are non-negative, and arranged in decreasing order, and $\mathbf p_i\in \R^n$ is the eigenvector corresponding to eigenvalue $\mu_i$.
Fixing $\varepsilon > 0$, we introduce the low-rank approximation
\begin{equation}
   \tilde B = \sum_{i: \mu_i > \varepsilon} \mu_i \mathbf p_i \mathbf p_i'.
\end{equation}
In other words, we treat the eigenvalues below $\varepsilon$ as 0. 
Let $k \le n$ denote the rank of $\tilde B$, i.e.~$k:=\sum_{i=1}^n 1\{\mu_i>\varepsilon\}$. 
Let $\mathbb Q$ be  the Potts model with $B$ replaced by $\tilde{B}$, i.e.
\begin{align}\label{eq:potts_low_rank}
  \mathbb Q(\mathbf X = \mathbf x) =\frac{1}{Z(\beta,\tilde B)} \exp \left (\frac{1}{2} \sum_{i, j = 1}^n \tilde{B}_{ij} 1 \{x_i = x_j \} \right).
\end{align}
The log unormalized probability mass $\mathbb Q$ can be written as
\begin{equation}\label{eq:low_rank_main}
  \frac{1}{2} \sum_{\ell= 1}^q \mathbf y_\ell' \tilde B \mathbf y_\ell
   = \frac{1}{2} \sum_{\ell= 1}^q \sum_{j= 1}^k \mu_j (\mathbf p_j' \mathbf y_\ell)^2.
\end{equation}
This representation motivates the following new auxiliary variable. Given $\mathbf X=\mathbf x$, let 
$\mathbf Z := \{Z_{\ell j}\}_{\ell \in [q], j \in [k]}$
be $kq$ mutually independent Gaussians with
\begin{align}\label{eq:main2}
Z_{\ell j} \sim \text{Normal}(\mathbf p'_j \mathbf y_\ell, 1 / \mu_j).
\end{align}
 Then the conditional distribution of ${\mathbf X}$ given ${\mathbf Z}$ is
\begin{equation}\label{eq:low_rank_conditional}
  \mathbb P(X_i = \ell \mid {\bf Z}) = \frac{\exp \left( \sum_{j = 1}^k \mu_j z_{\ell j} p_{ji} \right)}{\sum_{\ell'=1}^q \exp \left( \sum_{j = 1}^k \mu_j z_{\ell' j} p_{ji} \right)},
\end{equation}
which is straightforward to sample from. Also, the marginal density of ${\bf Z}$ has a similar low-rank structure and
\begin{align}\label{eq:low_rank_marginal}
p(\mathbf z) \propto 
\exp\Big(-\frac{1}{2}\sum_{\ell=1}^q \sum_{j=1}^k \mu_jz_{\ell j}^2\Big)\prod_{i=1}^n \Big(\sum_{\ell=1}^q \exp\Big(\sum_{j=1}^k\mu_jz_{\ell j} p_{ji} \Big)\Big)
\end{align}
The proof of both \eqref{eq:low_rank_conditional} and \eqref{eq:low_rank_marginal} are given in Appendix \ref{sec:choice:2}. 

By iterating between the steps \eqref{eq:main2} and \eqref{eq:low_rank_conditional}, we again have a block Gibbs sampler.
We no longer need to do an inversion and Cholesky decomposition but we do one Eigen decomposition instead, which incurs the same complexity $\mathcal O(n^3)$. The cost per iteration is dominated by (i) calculating the mean of each univariate normal, that is doing $qk$ inner-products of $n$-vectors in \eqref{eq:main2} and (ii) calculating the probability of sampling each state for each particle, which is $qn$ inner-products of $k$-vectors in \eqref{eq:low_rank_conditional}.
Both operations have complexity $\mathcal O(qnk)$.
The resulting Gibbs sampler with $m$ iterations has complexity $\mathcal O(n^3 + Tqnk)$.
This is an improvement over the previously obtained complexity $\mathcal O(n^3 + Tqn^2)$. In particular the improvement is significant when $T \gg n \gg k$.

A tuning parameter when approximating $A$ as low-rank is the threshold $\varepsilon$, under which small eigenvalues of the coupling matrix are replaced by 0.
To provide some guidance on setting $\varepsilon$, we bound the error on the low-rank AG sampler in the following proposition.
\begin{proposition}\label{thm:low-rank}
   Let $\mathbb P$ be the original Potts measure given in \eqref{eq:potts_simplified}, and let $\mathbb Q$ be the low-rank Potts measure given in \eqref{eq:potts_low_rank}. Then we have the following results:
\begin{itemize}
\item [] (i) The difference in the log normalizing constant of the two models satisfies

\[|\log Z(\beta,B)-\log Z(\beta,\tilde B)|\le \frac{1}{2} n\beta\varepsilon.\]

\item[]
(ii) With $\mathrm{KL}(.|.)$ denoting the Kullback-Leibler divergence between two probability measures, we have 
\[ \max\Big(\mathrm{KL}(\mathbb P| \mathbb Q), \mathrm{KL}(\mathbb Q| \mathbb P)\Big) \le n\beta\varepsilon.\]
\end{itemize}
\end{proposition} 
The proof is in Appendix \ref{sec:choice:3}.

In general, the low-rank auxiliary Gaussian is useful when $k={\rm rank}(\tilde{B})\ll n$. 
The following proposition provides a simple sufficient condition as to when this happens.
\begin{proposition}\label{prop:low_rank}
Suppose $\lambda_{\max}$ and $\lambda_{\min}$ are the largest and smallest eigenvalues of $A$. Let $B:=\beta(A+|\lambda_{\min}| I)$, and $\tilde{B}$ be its low-rank approximation for some $\varepsilon>0$. If 
\begin{align}\label{eq:mf}
\lambda_{\max}=O(1),\quad \lambda_{\min}=o(1),
\end{align}
 then $k:={\rm rank}(\tilde{B})=o(n)$.
%
\end{proposition}
The proof is in Appendix \ref{sec:choice:4}. 

The above lemma raises the question of whether there are natural examples of coupling matrices $A$ which satisfy \eqref{eq:mf}. Below we give some examples of matrices  $A$ which  satisfy \eqref{eq:mf}, and some examples which don't.
In cases where the Potts model is defined over a graph, we denote by $\tilde A$ the adjacency matrix---noting that $A \propto \tilde A$---and by $d$ the degree of a regular graph, that is the number of edges connected to each node.
\begin{itemize}
\item{\bf Satisfies \eqref{eq:mf}}:
Ising model on the Complete Graph (Curie-Weiss model), random regular graphs and Erd\H{o}s-R\'enyi graphs with large average degree, and certain Hopfield models:
\begin{enumerate}
\setlength{\itemsep}{1em}
    \item[(a)] {\bf Complete graphs.} Here, all the nodes are connected to one another and $A=\frac{1}{n}{\bf 1}{\bf 1}' - \frac{1}{n} I$.
    Hence $\lambda_{\max}=1-\frac{1}{n}=O(1)$ and $\lambda_{\min}=-\frac{1}{n}=o(1)$.
    \item[(b)] {\bf Random regular graphs.} For $d\leq n\slash 2$ and $A=\tilde{A}/d$, 
    we have $\lambda_{\max}(\tilde{A})=d$ and $\lambda_{\min}(\tilde{A})=O_P(\sqrt{d})$ (see \cite{cook2018size,tikhomirov2019spectral}). Consequently, as long as $d\rightarrow \infty$ with $n \to \infty$, we have $\lambda_{\max}=1=O(1)$ and $\lambda_{\min}=O_P\big(1 / \sqrt{d}\big)=o_P(1)$. 
    \item[(c)] {\bf Erd\"os--R\'enyi graphs.} For these graphs, we denote by $p$ the probability of there being an edge between two nodes.
    Then for $p= d/n$ and $A=\tilde{A} / d$, we have that when $d \gg \log n$, then $\big\|\tilde{A}-\frac{d}{n}{\bf 1}{\bf 1}'\big\|_2=O_P(\sqrt{d})$ \citep[Theorem 3.2]{benaych2020spectral}.
    Hence $\big\|A-\frac{1}{n}{\bf 1}{\bf 1}'\big\|_2= O_P\big(1 / \sqrt{d}\big)=o_P(1)$, and consequently $\lambda_{\max}=1+o_P(1)=O_P(1),\lambda_{\min}=o_P(1)$. 
    \item[(d)] {\bf Hopfield models.} The Hopfield model is defined by an $m \times n$ matrix $\boldsymbol \eta$ of i.i.d Rademacher random variables, i.e. $P(\eta_{ij} = \pm 1) = 1/2$.
    The coupling matrix is then $A = \text{off-diag}(\widetilde{A})$, where $\widetilde{A}=\boldsymbol \eta'\boldsymbol \eta / \max(m,n)$. Note that $A=\widetilde{A}-\frac{m}{\max(m,n)}I$. 

    When $n \ll m$, we have $\|\widetilde{A}-I\|_2=\big\|\frac{1}{m}\boldsymbol \eta'\boldsymbol \eta-I\big\|_2=O_P\big(\sqrt{\frac{n}{m}}\big)=o_P(1)$ (see \cite[Theorem 4.7.1]{vershynin2018high}). Consequently $\|A-{\bm 0}\|_2=o_P(1)$, and so $\lambda_{\max}=o_P(1), \lambda_{\min}=o_P(1)$.

    On the other hand, when $m\ll n$, we have $\widetilde{A}=\frac{1}{n}{\bm \eta}'{\bm \eta}$, and in this regime  \cite[Theorem 4.7.1]{vershynin2018high} gives
    $\frac{1}{n}{\bm\eta}{\bm \eta}'\stackrel{p}{\approx} I_m$. Since the nonzero eigenvalues of $\widetilde{A}=\frac{1}{n}{\bm \eta}'{\bm \eta}$ are the same as those of $\frac{1}{n}{\bm\eta}{\bm \eta}'\stackrel{p}{\approx} I_m$, we have $\lambda_{\max}(\widetilde{A})=1+o_P(1)$ and $\lambda_{\min}(\widetilde{A})=o_P(1)$. Consequently, $\lambda_{\max}=\lambda_{\max}(\widetilde{A})-\frac{m}{n}=1+o_P(1)$ and $\lambda_{\min}=\lambda_{\min}(\widetilde{A})-\frac{m}{n}=o_P(1)$.

    Note that the same applies if the components of $\boldsymbol \eta$ are $\sigma^2$-sub-Gaussian random variables for a fixed $\sigma$. Thus the only case where the Hopfield model is not approximately low-rank is the regime when $m\propto n$.
    


    %
\end{enumerate}




\item{\bf Does not satisfy \eqref{eq:mf}}:
 Ising model on graphs with bounded average (including Ising model on the $d$ dimensional integer lattice), Hopfield model with $m= n$, and Sherrington-Kirkpatrick model.

\end{itemize}
If \eqref{eq:mf} is not satisfied, then it seems reasonable to prefer the regular version of the auxiliary Gaussian algorithm over the low-rank version.
Numerical experiments suggest the regular version is faster when $A$ is full-rank.

\subsection{Continuous relaxation of the Potts model}


Rather than use a block Gibbs sampler which alternates between $({\bf Z} \mid {\bf X})$ and $({\bf X} \mid {\bf Z})$, it is possible to directly sample from $p({\bf Z})$ using \eqref{eq:marginal} (or \eqref{eq:low_rank_marginal}), which gives the unnormalized marginal density of the matrix ${\mathbf Z}$.
The discrete states are then recovered by sampling from $({\bf X} \mid {\bf Z})$.
This approach is particularly attractive, in that it lets us use gradient-based MCMC.
For example, \citet{Zhang:2012} use Hamiltonian Monte Carlo \citep[HMC;][]{Neal:2011, Betancourt:2018} to sample over the marginal distribution of ${\bf Z}$ for the Ising model.
Their strategy can be applied to Potts models and/or models with a low-rank coupling matrix $A$, using the auxiliary Gaussian variables we propose.

One difficulty is that $p({\bf Z})$ is highly multimodal, especially for large $\beta$, which can lead to long mixing times, even with gradient-based MCMC.
Furthermore, HMC is difficult to tune, and self-tuning strategies, such as the No-U Turn Sampler \citep[NUTS;][]{Hoffman:2014} are not straightforward to implement; a similar observation is made by \citet{Grathwohl:2021}.
In our numerical experiments, we consider an HMC algorithm using the proposed AG variables; however, the tuning parameters are held fixed and are likely suboptimal.
Finding a well-tuned gradient-based sampler for $p({\bf Z})$ is left as an open problem.

\section{Theoretical Analysis of Mixing Time} \label{sec:theory}

Monte Carlo estimators generated by MCMC suffer from a non-asymptotic bias, due to MCMC's biased initialization, i.e. $\mathbf x(0)$ is not drawn from \eqref{eq:potts}.
Hence, we need to overcome this initial bias during a \textit{burn-in} phase before accumulating samples to construct our Monte Carlo estimators. 
A common metric to quantify the number of necessary burn-in iterations is the total variation mixing time,
which can be used to bound the bias of Monte Carlo estimators over discrete spaces \citep[e.g.][]{Sinclair:1993, Diaconis:2009, MR3726904, Dwivedi:2019, Chen:2020}.
In this section, we present theoretical results on the mixing time of the AG sampler introduced in the previous section. The proofs of our theoretical results are deferred to Sections \ref{sec:proof1} and \ref{sec:proof2}. 

We first review total variation mixing time. 
Let $\mathbb N$ be the natural numbers and $\mathbb N^*$ the natural numbers excluding $0$.
Consider a Markov chain $X(t)$ with $t \in \mathbb N$ on a finite state space $\Omega$, and let $\pi$ denote the chain's stationary distribution.
For any $ x_0 \in \Omega$ and $t \in \mathbb N$, we denote by $K^t_{x_0}$ the distribution of $X(t)$ for a chain initialized at $X(0) = x_0$.
%
The total variation distance between $K_{x_0}^t$ 
and $\pi$ is defined as
\begin{equation*}
     \|K_{x_0}^t-\pi\|_{\mathrm{TV}}:=\frac{1}{2}\sum_{x\in\Omega}|K_{x_0}^t(x)-\pi(x)|, 
\end{equation*}
%
and takes values over $[0, 1]$.
%
Standard theory of finite Markov chains guarantees that the total variation distance converges to 0 as $t \to \infty$ under irreducibility and aperiodicity \citep[e.g.][]{Neal:1993,MR3726904}.
Unfortunately, this asymptotic result tells us little about the behavior of Markov chains over a finite number of iterations.

To better understand the pre-asymptotic behavior of MCMC, we study the $\epsilon$-mixing time.
Let $d(t):=\max_{x_0\in\Omega}\|K_{x_0}^t-\pi\|_{\mathrm{TV}}$. For any $\epsilon>0$, the $\epsilon$-mixing time is defined as
\begin{equation*}
    t_{\mathrm{mix}}(\epsilon):= \min \{t\in\mathbb{N}: d(t)\leq \epsilon\}.
\end{equation*}
Bounding $d(t)$ ensures that the result holds for any choice of initialization.

We now take $X(t)$ to be the $t^\text{th}$ sample generated by the AG sampler, and so $\Omega=[q]^n$ and $\pi$ is the probability mass function in \eqref{eq:potts}. 
To use the AG sampler, we need to construct a modified coupling matrix, $B = \beta (A + \lambda I)$,
for some $\lambda > |\lambda_\text{min}|$, where we recall that $\lambda_\text{min}$ is the smallest eigenvalue of $A$.
Any choice of $\lambda$ produces a correct Gibbs sampler, however for a small $\lambda$ we will obtain stronger guarantees on mixing time.
These guarantees are mediated by the 1-norm of $B$,
\begin{equation}
    \|B\|_1:=\max_{j\in [n]}\sum_{i=1}^n |B_{ij}|,
\end{equation}
which for a fixed $A$ is minimized for a small $\lambda$.




\begin{theorem}\label{Thm3.1}
Assume that $\|B\|_1< 4\slash q$. For any $\epsilon\in (0,1)$, the $\epsilon$-mixing time of the AG sampler satisfies
\begin{equation*}
    t_{\mathrm{mix}}(\epsilon)\leq \left\lceil \frac{\log{\epsilon^{-1}}+\log{n}}{-\log\big(q\|B\|_1\slash 4\big)} \right\rceil.
\end{equation*}
\end{theorem}

\begin{remark}
Theorem \ref{Thm3.1} applies to the Curie-Weiss model (Potts model on the complete graph, with the coupling matrix given by (\ref{coupling}); see Section \ref{subsubsec:CurieWeiss} for details), Potts models on lattice graphs (see Section \ref{subsubsec:Lattice}), and the Hopfield model:
\begin{itemize}
    \item For the Curie-Weiss model, we can take $\lambda$ slightly larger than $n^{-1}$. In this case, $\|B\|_1=\beta$ and our result implies that the AG sampler is fast mixing for $\beta\in (0,4\slash q)$. When $q=2$ (corresponding to the mean-field Ising model), this range covers the entire high temperature regime.
    
    \item For the Potts model on a lattice graph in $\mathbb{Z}^d$, we can take $A$ to be the adjacency matrix of the graph normalized by $2d$. In this case, $|\lambda_{\min}|\leq \|A\|_1\leq 1$ and any choice of $\lambda>1$ suffices. As $\|B\|_1\leq\beta(\|A\|_1+\lambda)\leq \beta(1+\lambda)$, our result implies that the AG sampler is fast mixing for $\beta\in (0,2\slash q)$. In fact, the same calculation shows that the AG sampler is fast mixing in the above regime when $A$ is the (scaled) adjacency matrix of \textit{any} $d_n$-regular graph, where $d_n$ can be fixed or grow with $n$ at an arbitrary rate.

    \item For the Hopfield model with $m\gg n^2\log n$, take $A=\text{off-diag}(\widetilde{A})$ with $\widetilde{A}=\boldsymbol \eta'\boldsymbol \eta / \max(m,n)$. Standard concentration bounds for sub-exponential random variables (see e.g. \cite[Corollary 2.8.3]{vershynin2018high}) and a union bound yield that 
    \begin{equation*}
        \|A\|_1=\frac{\max_{j\in[n]}\sum_{i\in [n]:i\neq j} \big|\sum_{k=1}^m \eta_{ki} \eta_{kj}\big|}{\max(m,n)}=\mathcal{O}\bigg(\frac{n\sqrt{m\log{n}}}{m}\bigg)=o(1)
    \end{equation*}
    with high probability. Hence the AG sampler is fast mixing.  
\end{itemize}
\end{remark}
\begin{remark}
The mixing time upper bound in Theorem \ref{Thm3.1} increases with $\|B\|_1$. As $\|B\|_1$ increases with $\lambda>|\lambda_{\text{min}}|$, our result provides a theoretical justification for choosing $\lambda$ to be only slightly larger than $|\lambda_{\min}|$.
This formalizes an empirical observation made by \citet{Martens:2010} that a small $\lambda$ yields better performance.
\end{remark}

Without the high temperature restriction, the mixing time can be at most exponential in $n\|B\|_1$, as shown in Theorem \ref{Thm3.2} below.


\begin{theorem}\label{Thm3.2}
For any $\epsilon\in (0,1)$, the $\epsilon$-mixing time of the AG sampler satisfies 
\begin{equation*}
    t_{\mathrm{mix}}(\epsilon)\leq \lceil \exp(2\|B\|_1 n) \log(\epsilon^{-1})\rceil.
\end{equation*} 
\end{theorem}
%
%
Theorem \ref{Thm3.2} gives a bound on the worst case mixing time, and it is very much possible to have faster mixing for particular models, even at low temperatures. 

To show this, we consider the low-rank AG sampler (Section~\ref{sec:lowrankAG}) with a permutation step (Section~\ref{sec:permutation}).
Theorem \ref{Thm3.3} shows that for the Curie-Weiss model (Potts model on the complete graph, with coupling matrix (\ref{coupling})), the AG sampler mixes within $\mathcal{O}(\log{n})$ iterations, even in the low temperature regime.


\begin{theorem}\label{Thm3.3}
Consider the Curie-Weiss model with $q\geq 2$, and assume that the inverse temperature $\beta>q$. For any $\epsilon\in (0,1)$, the $\epsilon$-mixing time of the modified AG sampler described as above satisfies
\begin{equation*}
    t_{\mathrm{mix}}(\epsilon) \leq C  \log(n+1) \lceil\log(\epsilon^{-1}) \rceil,
\end{equation*}
where $C$ only depends on $q,\beta$. 
\end{theorem}
\begin{remark}
For the Curie-Weiss model with general $q$, there are three critical temperatures $\beta_u\leq \beta_o\leq\beta_{rc}=q$. Following \cite{Galanis:2019}, $\beta_u$ corresponds to the uniqueness\slash non-uniqueness threshold, $\beta_o$ corresponds to the ordered\slash disordered phase transition, and $\beta_{rc}$ was conjectured to correspond to a second uniqueness\slash non-uniqueness threshold (\cite{Haggstrom:1996}); see \cite{Costeniuc:2005, Cuff:2012, Galanis:2016} for further details. When $\beta\in (\beta_u,\beta_{rc})$, the Curie-Weiss model exhibits metastable behavior and we expect common MCMC samplers to be slow mixing in this regime (see e.g. \cite{Galanis:2019,coja2023metastability}). Theorem \ref{Thm3.3} implies fast mixing of the modified AG sampler for the Curie-Weiss model throughout the low temperature regime $\beta>\beta_{rc}$.
\end{remark}

\section{Numerical Experiments} \label{sec:experiment}

In this section, we numerically evaluate the performance of the AG sampler across a range of Ising and Potts models.
The code to reproduce the experiments and figures can be found at \url{https://github.com/charlesm93/potts_simulation}.

As benchmarks, we consider several MCMC algorithms.
An important class of algorithms are Metropolis algorithms, which propose a new state ${\bf x}'$ and accept it with a certain probability.
Specifically, we consider Barker's acceptance rule, which is to accept the proposed state with probability,
\begin{equation}
    \alpha = \frac{\mathbb P({\bf x}')}{\mathbb P({\bf x}) + \mathbb P({\bf x}')} = \frac{1}{1 + e^{\mathcal L({\bf x}) - \mathcal L({\bf x}')}},
\end{equation}
where $\mathcal L({\bf x}) := \log \mathbb P({\bf x})$.
In the special case where $q=2$ and the proposal is obtained by ``flipping'' a single element $x_i$, an iteration of the Metropolis algorithm with Barker's acceptance rule is exactly one iteration of the heat bath algorithm or Gibbs sampler, which draws a new state from the conditional distribution $\mathbb P(X_i \mid {\bf x}_{-i})$.
A crucial choice in the Metropolis algorithm is to determine which site $i$ to update.
This can be done by sampling $i$ uniformly from $[n]$ at each iteration, using a schedule that sequentially updates each site, or with a more sophisticated proposal as in Gibbs-with-Gradient \citep{Grathwohl:2021}.
The calculation of $\mathcal L({\bf x}) - \mathcal L({\bf x}')$ can be done by treating $\mathcal L({\bf x})$ as a user-provided black box function with computational cost $\mathcal O(n^2)$.
However, if the difference between ${\bf x}$ and ${\bf x}'$ is only one element, then $\mathcal L({\bf x}) - \mathcal L({\bf x}')$ can be computed in $\mathcal O(n)$ operations, given an initial calculation of $\mathbb P({\bf x})$ at the first iteration of MCMC.
All the Metropolis algorithms we consider update a single site and so, unless otherwise specified, we use the specialized calculation of $\mathcal L({\bf x}) - \mathcal L({\bf x}')$.

We also consider the Wolff algorithm \citep{Wolff:1989} for Potts models defined on graphs.
The Wolff algorithm is similar to the Swendsen-Wang algorithm but only updates a single cluster per iteration.
In preliminary runs, we found that the Wolff algorithm consistently outperforms the Swendsen-Wang algorithm and so we only report results for the Wolff algorithm.

Finally, we examine three samplers which use an auxiliary Gaussian variable: the AG sampler, the AG sampler with a low-rank auxiliary Gaussian, and discrete Hamiltonian Monte Carlo. The algorithms we evaluate are summarized in Table~\ref{tab:algorithms}.

\begin{table}
    \centering
    \renewcommand{\arraystretch}{1.25}
    \begin{tabular}{l l}
         \rowcolor{Gray} {\bf Algorithm} & {\bf Description}  \\
         Black Box Metropolis & Random site update with black box calculation of $\mathcal L ({\bf x}) - \mathcal L({\bf x}')$. \\
         \rowcolor{Gray} Metropolis & Random single site update. \\
         Metropolis Long & Single site update from sequence. \\
         \rowcolor{Gray} Gibbs with Gradient & Single site update using gradient-based proposal. \\
         Wolff & Multiple updates using an auxiliary cluster. Only works for Potts\\
         & models defined on a graph. \\
         \rowcolor{Gray} AG & Alternates between updating ${\bf X}$ and an AG variable ${\bf Z}$ \\
         \rowcolor{Gray} & (Section~\ref{sec:PottsAG}). When $q = 2$, the AG variable is specialized and we \\
         \rowcolor{Gray} & recover the algorithm by \citet{Martens:2010}. \\
         AG Low-Rank & Same as above, but with low-rank AG variable (Section~\ref{sec:lowrankAG}). \\
         \rowcolor{Gray} Discrete HMC & Static Hamiltonian Monte Carlo over the space of the AG \\
         \rowcolor{Gray} & variable ${\bf Z}$ (Section~\ref{sec:PottsAG}). 
    \end{tabular}
    \caption{Summary of MCMC algorithms in numerical experiments.
    \textit{The Metropolis, Metropolis Long, and Gibbs with Gradient use a specialized calculation of the acceptance probability.}
    }
    \label{tab:algorithms}
\end{table}

\subsection{Performance metric}

A typical goal with MCMC is to estimate expectation values, that is, for a function $\phi: [q]^n \to \Phi \subseteq \mathbb R$, we estimate $\mathbb E \left [\phi(\mathbf X) \right]$.
Our estimator is
\begin{equation}
    \hat \phi_T = \frac{1}{\ceil{T/2}} \sum^T_{t = \ceil{T/2}} \phi \left ( \mathbf x^{(t)} \right),
\end{equation}
where $(\mathbf x(0), \cdots, \mathbf x(T))$ are samples obtained via MCMC and $T$ is the total number of iterations.
Following \citet{Gelman:2011}, we discard the first half of the Markov chain as part of a burn-in phase.
As a quantity of interest, we focus on \mbox{$\phi({\bf x}) = - \sum_{i,j = 1}^n A_{ij} 1 \{x_i = x_j \}$}, which is the system's Hamiltonian and is a sufficient statistics of \eqref{eq:potts}.

We compute two statistics to evaluate the quality of $\hat \phi_T$:
\begin{itemize}

\item[(i)] The $\widehat R$-statistics \citep{Gelman:1992, Vehtari:2020}, which compares the mean of multiple chains initialized at distinct points and approaches 1 as the Markov chain converges.

\item[(ii)] The effective sample size (ESS), which for $T$ large enough can be interpreted as the number of independent samples required to achieve the variance of $\hat \phi_T$ \citep{Roberts:2004}.
Following the recommendation of \citet{Vehtari:2020}, we will use the ESS to characterize the error of $\hat \phi_T$, provided $\widehat R \approx 1$.
\end{itemize}
We compute both quantities using the R package {\sc Posterior} \citep{posterior}.

Our goal is to achieve convergence and a target ESS in as short a time as possible.
There is no basic operations common to all the algorithms we consider, so the computational cost of a sampler is measured by its runtime (wall time).
This means that our results depend on the implementation details of each algorithm.
For this experiment, all samplers are implemented in R. 
%
For each algorithm, we run 4 chains in order to compute $\widehat R$.
For the AG sampler, each chain is made of 50,000 iterations.
The length of the chains for other algorithms is roughly set to keep total runtimes between methods comparable.

\subsection{Potts models on a graph}

We begin with the case where $A$ is the adjacency matrix of a graph $G$ scaled by its average degree. 
We consider two choices of $G$ which are of wide interest in statistical physics: the Ising model on the integer lattice and the Curie-Weiss model.
%

\subsubsection{Lattice graph}\label{subsubsec:Lattice}

\begin{figure}
    \centering
    \includegraphics[width=6in]{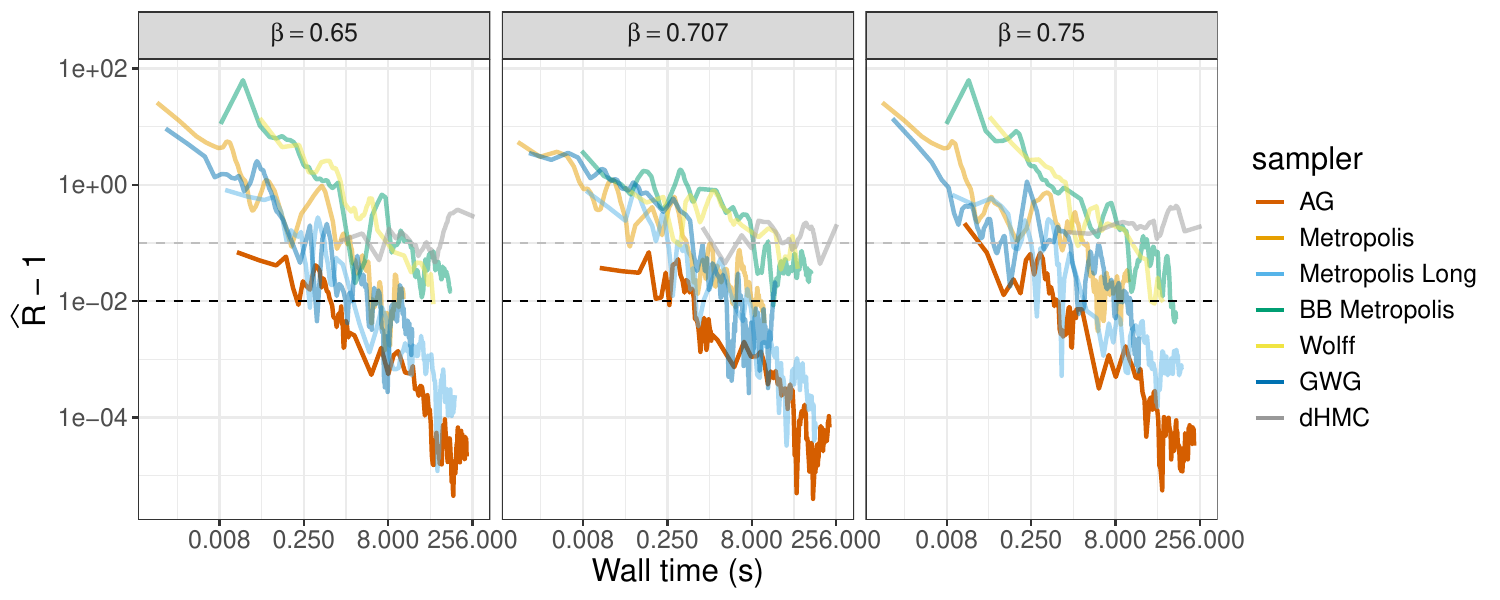}
    \caption{$\widehat R$ over time. \textit{Results on a $24 \times 24$ Ising grid at or near the critical temperature $\beta_c = 1 / \sqrt{2} \approx 0.707$.
    At various checkpoints, we evaluate the run time and $\widehat R$ for each sampler.
    $\widehat R \le 1.1$ indicates the sampler is nearing convergence; in practice the more conservative threshold $\widehat R \le 1.01$ is recommended.
    Overall, the AG sampler achieves the fastest convergence, as measured by $\widehat R$. 
    }}
    \label{fig:rhat}
\end{figure}

\begin{figure}
    \centering
    \includegraphics[width=5in]{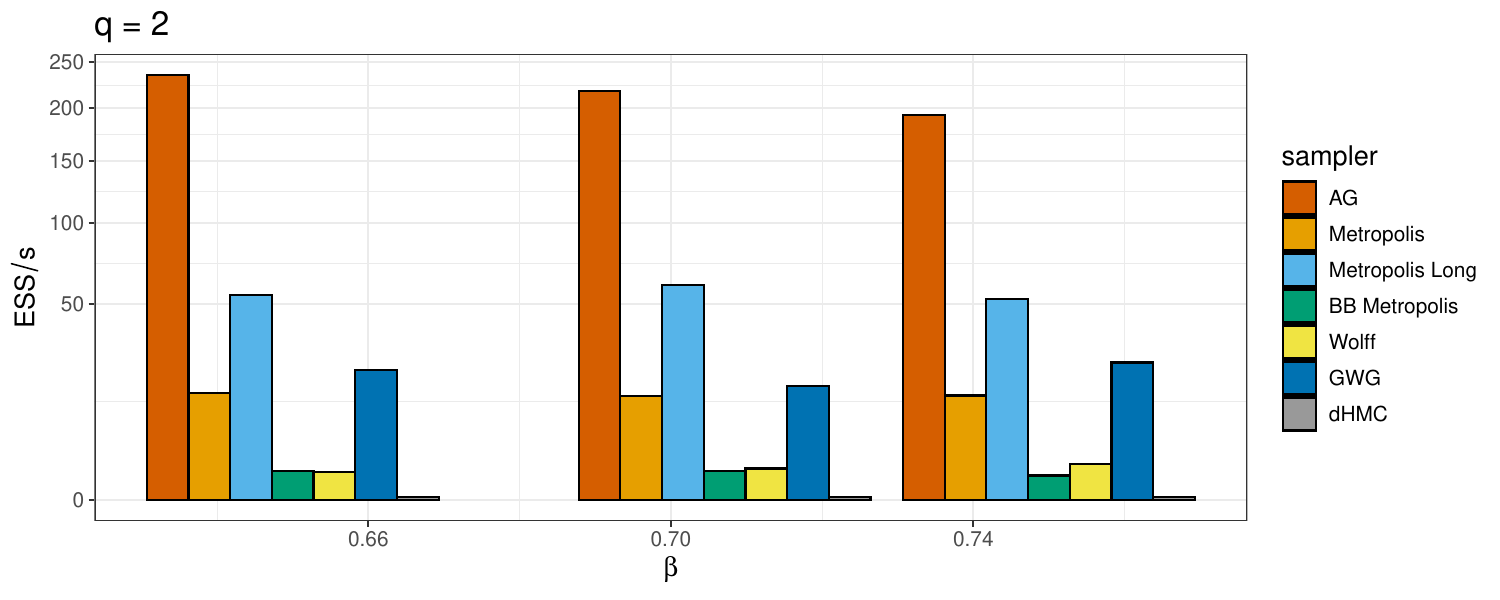}
    \includegraphics[width=5in]{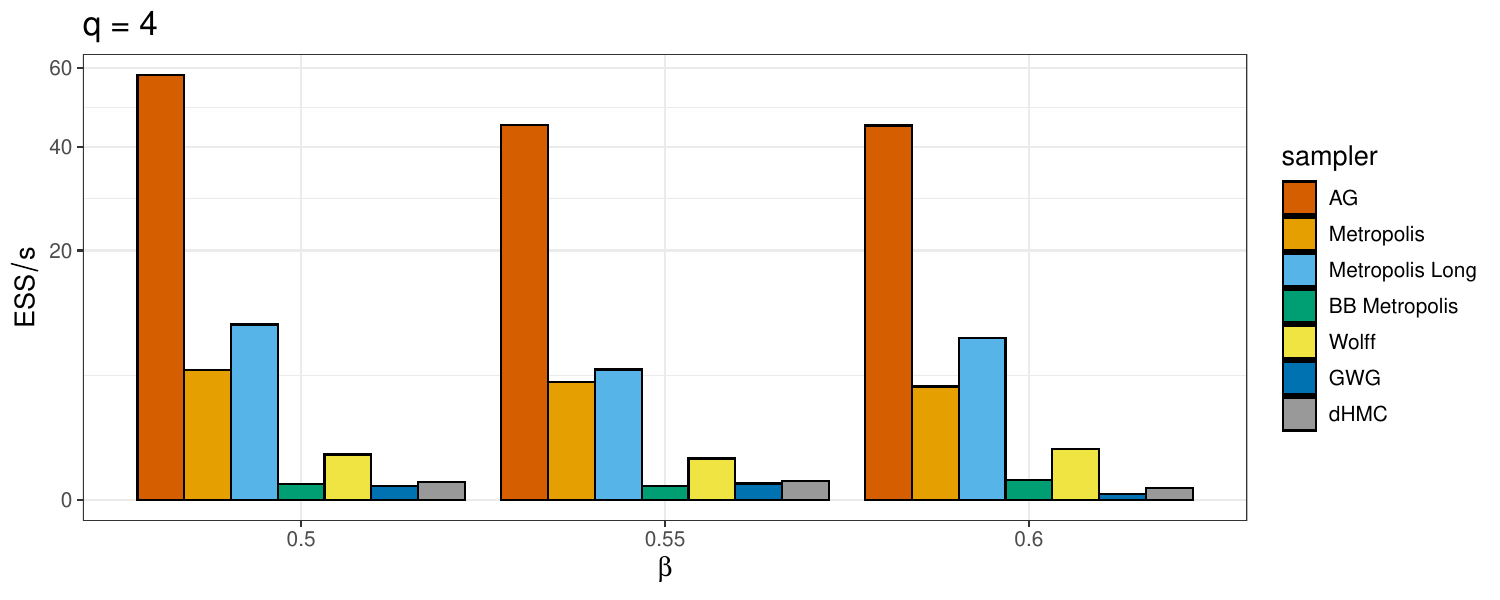}
    \caption{Effective sample size per second on a $24 \times 24$ grid.
    \textit{(Above) Ising model, i.e. $q = 2$, and the critical temperature is $\beta_c = 1 / \sqrt{2}$.
    (Below) Potts model with $q = 4$ and the critical temperature is $\beta_c = 0.55$.
    }}
    \label{fig:ess_ising_grid}
\end{figure}

The two-dimensional lattice graph with two states (i.e.~\mbox{$q = 2$}) is possibly the best studied example. 
We examine a $24 \times 24$ graph (i.e. $n=576$), and consider the cases $q = 2, 4$.
We focus on a parameter range in $\beta$ around the critical temperature $\beta_c(q)$, which is approximately $0.71$ and $0.55$ for $q=2$ and $q=4$ respectively, as reported by \cite{Monroe:2002}.
For $q = 2$, we illustrate the use of $\widehat R$ (Figure~\ref{fig:rhat}).
We find that all algorithms achieve $\widehat R \lesssim 1.1$, which suggests approximate convergence (a more conservative threshold would be 1.01), and so calculations of the ESS provide a reasonable characterization of the Monte Carlo estimator's error.
We find that $\widehat R$ decays fastest for the AG sampler.
On the other hand, HMC struggles to converge and we attribute this to the difficulty of finding suitable tuning parameters which work well when sampling from $p({\bf Z})$.

Figure~\ref{fig:ess_ising_grid} plots the ESS/s for $q=2$ and $q=4$.
In both cases, the AG sampler offers the best performance.
The results also highlight important variations between different Metropolis algorithms.
The black box Metropolis performs poorly but we see a net improvement when using a specialized calculation of $\mathcal L({\bf x}') - \mathcal L({\bf x})$.
Gibbs-with-gradient performs well for $q=2$ but the performance decreases for $q=4$;
we attribute this to the increased cost of computing a gradient-based proposal for Potts models.
Overall Metropolis Long, which updates the sites in sequence, produces the best performance among Metropolis algorithms.
More sophisticated strategies, such as the Wolff algorithm provide better ESS per iteration but this does not justify the high wall time cost per iteration.
We highlight that our implementation of Wolff does not exploit the sparsity of $A$ and that discrete HMC used static (and likely suboptimal) tuning parameters.\footnote{Pilot runs with the self-tuning HMC implemented in {\sc Stan} \citep{Carpenter:2017} also did not produce convincing results.}

\subsubsection{Curie-Weiss model}\label{subsubsec:CurieWeiss}

In this model $A_n={\bf 1}{\bf 1}' / n - I_n / n$ is the scaled adjacency matrix of a complete graph, where each node is connected to every other node.
Once again, we consider the cases $q = 2, 4$ with $n = 576$.
The critical value $\beta_c(q)$ is  $1$ and $\sim$1.64 for $q=2$ and $q=4$ respectively  \citep[e.g][]{Bollobas:1996}, around which we focus our simulations.
Adjusting the diagonal elements, we obtain a rank-1 coupling matrix on which we can apply the low-rank AG variable.

\begin{figure}
    \begin{center}
        \includegraphics[width=5in]{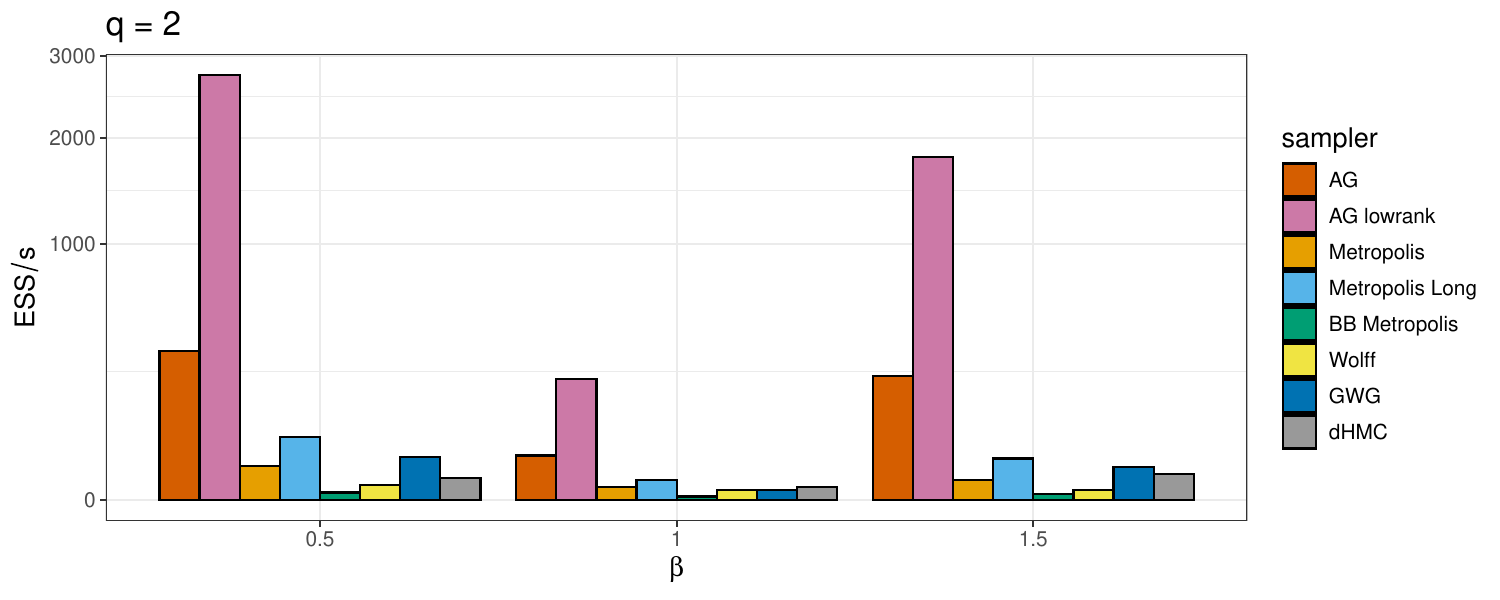}
        \includegraphics[width=5in]{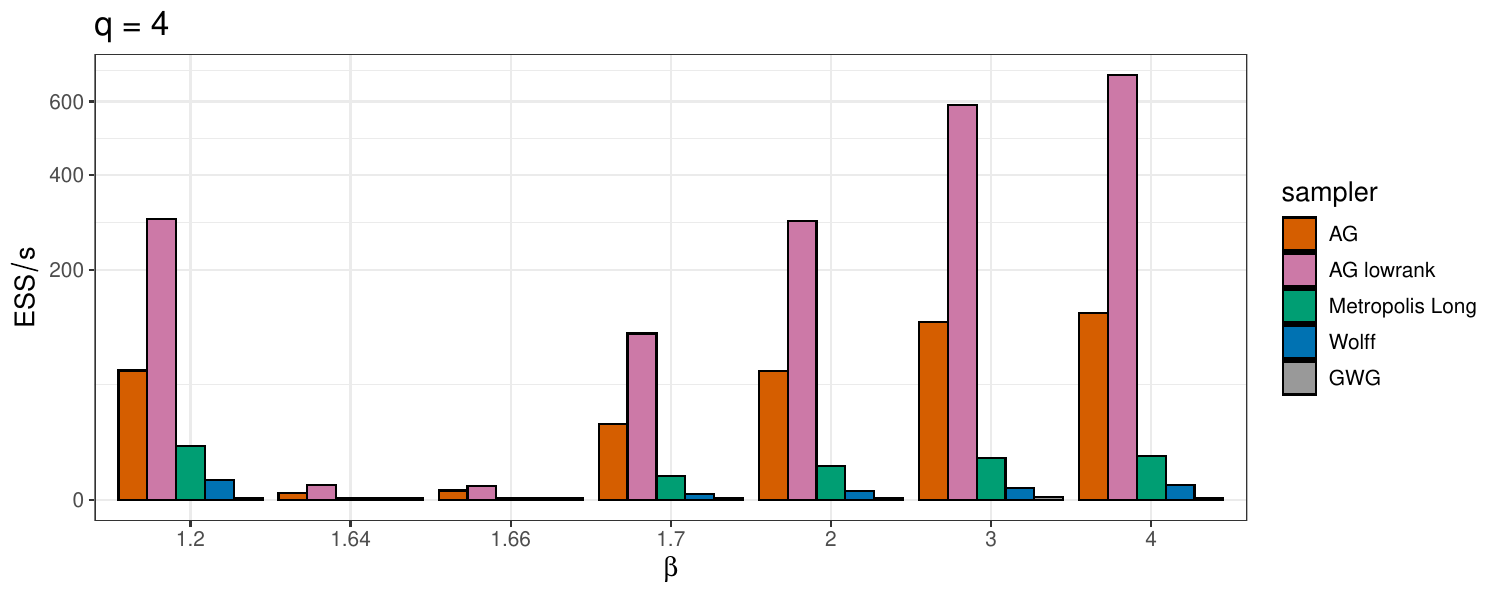}
    \end{center}
    \caption{Effective sample size per second on a $24 \times 24$ fully-connected graph.
    \textit{(Above) Ising model, i.e. $q = 2$. The critical temperature occurs at $\beta_c = 1$ and incurs a decrease in efficiency for all algorithms.
    Using a low-rank AG improves performance.
    (Below) For $q = 4$, we observe  a critical slow down for all algorithms near the critical temperature $\beta_c \sim 1.65$.
    We also obtain competitive performance for a large $\beta$ (cold temperatures). 
    }}
    \label{fig:ess_complete}
\end{figure}

For $q = 2$, the results are consistent with what we observed on the grid model, with the AG sampler producing the most competitive results (Figure~\ref{fig:ess_complete}).
Using a low-rank AG offers a dramatic improvement, as might be expected, since for a fully-connected graph, the dimension of the auxiliary Gaussian variables can be reduced from $n = 576$ to $k = 1$.
All algorithms suffer from an observable slow-down at the critical temperature, $\beta_c = 1$.

Similarly for $q=4$ (Figure~\ref{fig:ess_complete}), we observe a critical slow down near $\beta \approx 1.65$.
Away from this point, the performance of all samplers improves, including at cold temperatures ($\beta \ge q$).
Theorems~\ref{Thm3.1} and \ref{Thm3.3} predict fast mixing times for the AG algorithms when $\beta \notin (1, 4)$.
The experiment reveals that AG samplers can also perform well within $\beta\in (1,4)$, and that the exponential bound by Theorem~\ref{Thm3.2} is conservative.
Finally, we note that the critical slow down around $\beta \approx 1.65$ has been analyzed in previous studies on the heat bath \citep{Cuff:2012};
empirically we find this result to generalize to AG samplers.

\subsection{Spin glass models}\label{section:spin}

In a spin glass model, the entries in $A$ can take both positive and negative values.
Two important examples are the Hopfield model and the Sherrington-Kirkpatrick (SK) model. For these problems, auxiliary cluster algorithms, including the Wolff and Swendsen-Wang algorithms, are no longer an option.
Moving forward, we only use Metropolis Long (with sequential updates) as a benchmark.

\subsubsection{Hopfield Model}


Let \mbox{$\boldsymbol\eta \in \mathbb R^{m \times n}$} be a matrix of i.i.d random variables with \mbox{$P(\eta_{ik} = \pm 1) = 1 / 2$},
and let 
\begin{align}\label{eq:hop}
A=\frac{\boldsymbol\eta' \boldsymbol\eta}{\max(m,n)}.
\end{align}
$A$ can take values which are positive, negative, or zero, meaning the spins $X_i$ can be correlated,  anti-correlated, or uncorrelated.



\begin{figure}
    \centering
    \includegraphics[width=4.5in]{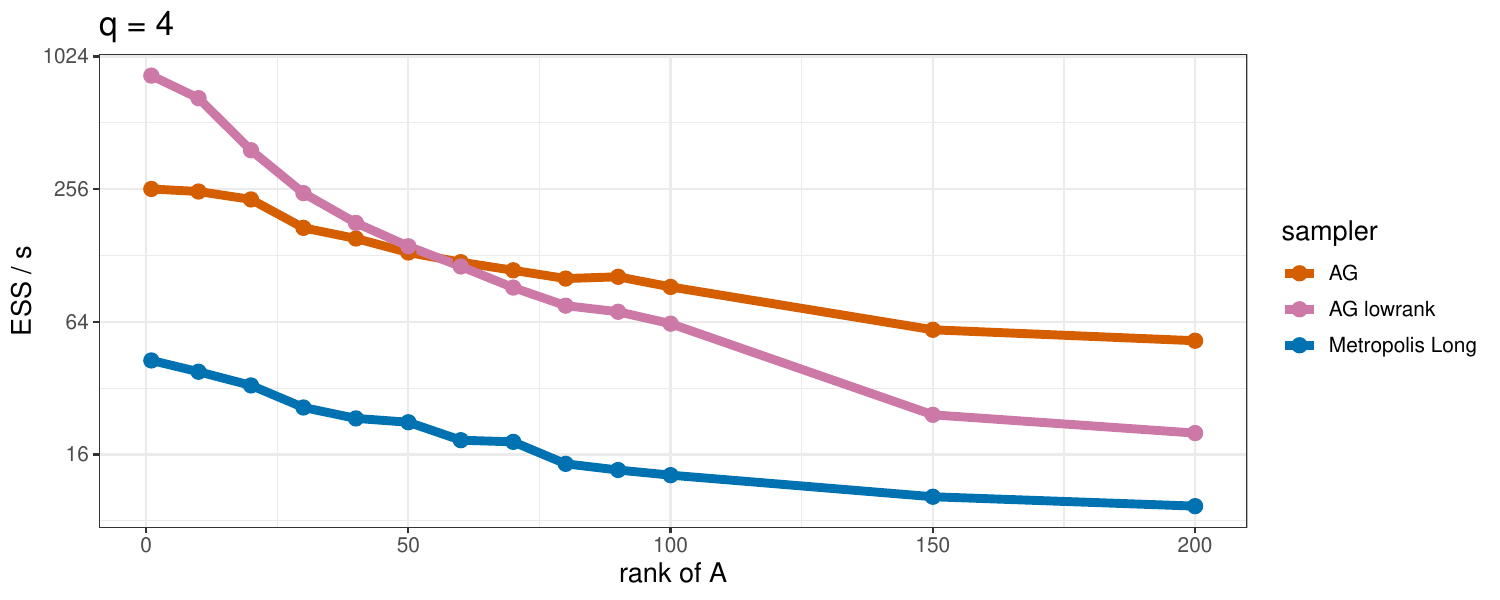}
    \caption{Effective sample size per second on a $24 \times 24$ Hopfield model. \textit{For a low-rank coupling matrix, the low-rank AG offers a large improvement in performance. For larger ranks, the regular AG sampler performs slightly better. For higher ranks, the performance of all algorithms decreases.}}
    \label{fig:hopfield}
\end{figure}

We use this example to examine the validity of the above prediction, and to compare the regular and the low-rank AG samplers across varying ranks. 
We set $\beta = 1$, $q = 4$, $n = 576$, and vary $m$ from 1 to 200.
We plot the results in Figure~\ref{fig:hopfield}, and find AG samplers to outperform the Metropolis sampler, with the low-rank AG yielding a large improvement for small $m$.

\begin{remark}
In all our experiments, the regular and low-rank AG generate roughly the same effective sample size.
The difference in performance is therefore driven by the difference in the computational cost of each iteration. 
\end{remark}

\subsubsection{Cold SK model}

In the SK model, the elements of a coupling matrix $A_{ij}$ are mutually independent, drawn from a standard normal scaled by $1 / \sqrt n$, equivalently a normal with variance $1 / n$:
\begin{equation*}
  A_{ij} = A_{ji} \sim \text{Normal} \left (0, n^{-1} \right).
\end{equation*}
This coupling matrix unfortunately does not admit a good low-rank approximation (c.f.~\citet[Sec 1.3.2]{Basak:2017}).
An important application of this model in physics is the study of spin glass systems at cold temperatures \citep[e.g][]{Yucesoy:2013, Katzgraber:2001}.
For high $\beta$, the Potts model becomes highly multimodal, resulting in a slow or incomplete exploration by the Markov chains.  

Parallel tempering can mitigate this problem \citep{Geyer:1991, Hukushima:1998}; see also \citet{Swendsen:1986} for an earlier related proposal.
Parallel tempering runs multiple chains or \textit{replicas} over a schedule of temperatures for a set number of iterations and then exchanges, with a certain probability, the states of two replicas with a ``neighboring'' temperature.
The probability of exchanging two neighboring replicas is chosen so as to maintain detailed balance.
At high temperatures, the Markov chain can move more easily across the target space and overcome the energy barriers between modes, before being cooled down again to sample at the temperature of interest.

When implementing a tempering algorithm, we need to choose the number of replicas and their respective temperatures.
The first replica is at the cold temperature of interest.
We then progressively increase the temperature of each replica to ensure that the exchange probability is high and that we eventually reach a warm temperature at which the sampler is not frustrated by high energy barriers.
In our experiment, the target inverse-temperature is $\beta= 3$. 
We create 11 replicas at evenly spaced temperatures, \mbox{$\beta = (0.5, 0.75, 1, 1.25, 1.5, 1.75, 2, 2.25, 2.5, 2.75, 3)$}.
This naive approach yields reasonable, if somewhat uneven, acceptance rates for the exchange proposals.
There is extensive literature on setting the temperature schedule for parallel tempering using more principled, albeit more intricate, approaches \citep[e.g.][]{Syed:2022, Atchade:2011, Kone:2005, Hukushima:1998} and we leave the analysis of such tempering strategies with AG sampler to future work.

In total, we attempt 40 exchanges and run 2,000 iterations between exchanges, for a total of 80,000 iterations.
We run 4 sets of replicas with different initializations to compute $\widehat R$ and monitor whether the chains are mixing. 
The size of the system is $n = 128$.

\begin{figure}
    \centering
    \includegraphics[width=5.5in]{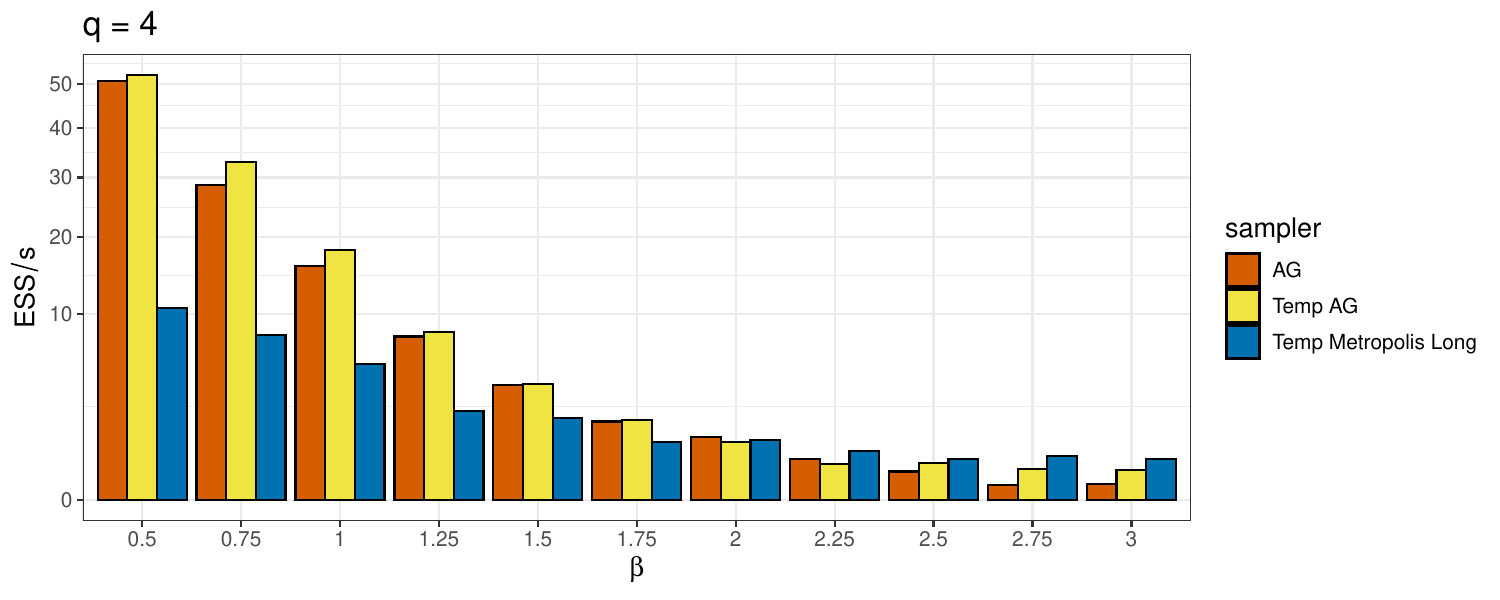}
    \caption{Effective sample size per second for a Spin Glass system with $n = 128$. \textit{At cold temperatures (high $\beta$), the tempered scheme improves over the non-tempered AG sampler.
    However the advantage of the AG sampler over the Heat Bath vanishes for $\beta \ge 2$.
    }}
    \label{fig:sk}
\end{figure}

We plot the results of our experiment in Figure~\ref{fig:sk}.
The non-tempered AG suffers at cold temperatures, notably for $\beta \ge 2.5$.
The tempered AG is competitive across all temperature regimes, matching the superiority of the regular AG sampler for small $\beta$s, and yielding reasonable results for $\beta$ large.
However, we did not find that AG methods outperform the tempered Metropolis  sampler for $\beta \ge 2$.

\section{Proofs of Theorems \ref{Thm3.1} and \ref{Thm3.2}}\label{sec:proof1}

In this section, we present the proofs of Theorems \ref{Thm3.1} and \ref{Thm3.2} in Sections \ref{subsec:proof:Thm3.1} and \ref{subsec:proof:Thm3.2}, respectively. For any $\mathbf{x}=(x_1,\cdots,x_n),\mathbf{x}'=(x_1',\cdots,x_n')\in [q]^n$, we denote the Hamming distance between $\mathbf{x}$ and $\mathbf{x}'$ by $H(\mathbf{x},\mathbf{x}'):=\sum_{i=1}^n 1\{x_i\neq x_i'\}$. 

\subsection{Proof of Theorem \ref{Thm3.1}}\label{subsec:proof:Thm3.1}

The proof of Theorem \ref{Thm3.1} is based on the path coupling technique \cite[Chapter 14]{MR3726904}. To apply this technique, we associate a labeled undirected graph with vertex set $[q]^n$, so that any two states $\mathbf{x}=(x_1,\cdots,x_n),\mathbf{x}'=(x_1',\cdots,x_n')\in [q]^n$ are connected by an edge if and only if $H(\mathbf{x},\mathbf{x}')=1$.

In the following, we consider any two distinct states $\mathbf{x}=(x_1,\cdots,x_n),\mathbf{x}'=(x_1',\cdots,x_n')\in [q]^n$. Denote by $\mathbf{X}(1)$ and $\mathbf{X}'(1)$ the new states after one iteration of the AG sampler started from $\mathbf{x}$ and $\mathbf{x}'$, respectively. We aim to construct a coupling of $\mathbf{X}(1)$ and $\mathbf{X}'(1)$ such that $\mathbb{E}[H(\mathbf{X}(1),\mathbf{X}'(1))]$ is small.

We first rewrite the steps taken by the AG sampler.
For each $\ell\in [q]$, let 
$$\mathbf{y}_{\ell} =(1\{x_1=\ell\},\cdots,1\{x_n=\ell\})',\qquad \mathbf{y}'_{\ell} =(1\{x_1' =\ell\},\cdots,1\{x_n' =\ell\})'\in\mathbb{R}^n.$$
Let $\{\mathbf{w}_{\ell}\}_{\ell\in [q]} \stackrel{i.i.d.}{\sim} N(0,B^{-1})$. 
For each $\ell\in [q]$, let 
$$\mathbf{z}_{\ell}=\mathbf{y}_{\ell}+{\bf w}_{\ell},\qquad \mathbf{z}'_{\ell}=\mathbf{y}'_{\ell}+ {\bf w}_{\ell}.
$$ 
%
Given $\mathbf{Z}=\{\mathbf{z}_{\ell}\}_{\ell=1}^q$ and $\mathbf{Z}'=\{\mathbf{z}'_{\ell}\}_{\ell=1}^q$, we independently sample the pairs $\{(X_{i}(1),X_{i}'(1))\}_{i=1}^n$, so that for all $i\in [n]$ and $\ell\in [q]$,
\begin{equation}\label{Eq5.8}
    \mathbb{P}(X_{i}(1)=\ell\mid \mathbf{Z}, \mathbf{Z}')=\frac{ \exp \left ( \sum_{j = 1}^n B_{ij} z_{\ell j} \right )}{\sum_{\ell'=1}^q\exp \left ( \sum_{j = 1}^n B_{ij} z_{\ell' j} \right )},
\end{equation}
\begin{equation}\label{Eq5.9}
    \mathbb{P}(X'_{i}(1)=\ell\mid \mathbf{Z}, \mathbf{Z}')=\frac{ \exp \left ( \sum_{j = 1}^n B_{ij} z'_{\ell j} \right )}{\sum_{\ell'=1}^q\exp \left ( \sum_{j = 1}^n B_{ij} z'_{\ell' j} \right )},
\end{equation}
\begin{eqnarray}\label{Eq5.1}
  &&  \mathbb{P}(X_{i}(1)\neq X_{i}'(1) |\mathbf{Z}, \mathbf{Z}')=\|\mathcal{L}_{X_{i}(1)\mid \mathbf{Z}, \mathbf{Z}'}-\mathcal{L}_{X'_{i}(1)\mid \mathbf{Z}, \mathbf{Z}'}\|_{\mathrm{TV}}\nonumber\\
  &=&\frac{1}{2}\sum_{\ell=1}^q\Bigg|
  \frac{ \exp \left ( \sum_{j = 1}^n B_{ij} z_{\ell j} \right )}{\sum_{\ell'=1}^q\exp \left ( \sum_{j = 1}^n B_{ij} z_{\ell' j} \right )}
  -
  \frac{ \exp \left ( \sum_{j = 1}^n B_{ij} z'_{\ell j} \right )}{\sum_{\ell'=1}^q\exp \left ( \sum_{j = 1}^n B_{ij} z'_{\ell' j} \right )}
  \Bigg|,
\end{eqnarray}
where $\mathcal{L}_{X_{i}(1)\mid \mathbf{Z}, \mathbf{Z}'}$ and $\mathcal{L}_{X'_{i}(1)\mid \mathbf{Z}, \mathbf{Z}'}$ are the conditional distributions of $X_{i}(1)$ and $X'_{i}(1)$ given $\mathbf{Z}, \mathbf{Z}'$. Note that this can be done due to \cite[Proposition 4.7]{MR3726904}. 

For each $\ell\in [q]$, we define $T_{\ell}: \mathbb{R}^q\rightarrow (0,1)$ by
\begin{equation}\label{deftl}
    T_{\ell}(m_1,\cdots,m_q):=\frac{\exp(m_{\ell})}{\sum_{\ell'=1}^q \exp(m_{\ell'})},\text{ for any }(m_1,\cdots,m_q)\in\mathbb{R}^q.
\end{equation}
Then for any $r\in [q]$,
\begin{equation*}
  \frac{\partial T_{\ell}(m_1,\cdots,m_q)}{\partial m_r}= T_{\ell}(m_1,\cdots,m_q)(\mathbbm{1}_{\ell=r}-T_r(m_1,\cdots,m_q)).   
\end{equation*}
Hence, for $r = \ell$,
\begin{equation}\label{Eq5.2}
    \bigg|\frac{\partial T_{\ell}(m_1,\cdots,m_q)}{\partial m_{\ell}}\bigg|=T_{\ell}(m_1,\cdots,m_q)(1-T_{\ell}(m_1,\cdots,m_q))\leq \frac{1}{4},
\end{equation}
and for any $r \neq \ell$,
\begin{eqnarray}\label{Eq5.3}
    \bigg|\frac{\partial T_{\ell}(m_1,\cdots,m_q)}{\partial m_{r}}\bigg|&=&T_{\ell}(m_1,\cdots,m_q)T_r(m_1,\cdots,m_q)\nonumber\\
    &\leq&T_{\ell}(m_1,\cdots,m_q)(1-T_{\ell}(m_1,\cdots,m_q))\leq \frac{1}{4},
\end{eqnarray}
where the first inequality above follows from
\begin{eqnarray*}
    T_r(m_1, \cdots, m_q) = \frac{\exp(m_r)}{\sum_{\ell'=1}^q \exp(m_{\ell'})}
     &\le&  \sum_{r' \neq \ell} \frac{\exp(m_{r'})}{\sum_{\ell'=1}^q \exp(m_{\ell'})} \\
    & = & 1 - \frac{\exp(m_\ell)}{\sum_{\ell'=1}^q \exp(m_{\ell'})} = 1 - T_\ell(m_1, \cdots, m_q).
\end{eqnarray*}
For each $i\in [n]$ and $\ell\in [q]$, we define
\begin{equation*}
    M_{i\ell}:=\sum_{j = 1}^n B_{ij} z_{\ell j}, \quad  M'_{i\ell}:=\sum_{j = 1}^n B_{ij} z'_{\ell j}.
\end{equation*}
Note that
\begin{equation}\label{Eq5.4}
    |M_{i\ell} - M'_{i\ell}|\leq \sum_{j=1}^n |B_{ij}||z_{\ell j}-z_{\ell j}'|=\sum_{j=1}^n |B_{ij}||1\{x_j=\ell\}-1\{x_j'=\ell\}|.
\end{equation}
We now bound $\mathbb{E}[H(\mathbf{X}(1),\mathbf{X}'(1))]$ as follows. By (\ref{Eq5.1})-(\ref{Eq5.4}), 
\begin{eqnarray}
  && \mathbb{E}[H(\mathbf{X}(1),\mathbf{X}'(1))|\mathbf{Z},\mathbf{Z}']
  =\sum_{i=1}^n\mathbb{P}(X_{i}(1)\neq X'_{i}(1)|\mathbf{Z},\mathbf{Z}') \nonumber\\
  &=& \frac{1}{2}\sum_{i=1}^n\sum_{\ell=1}^q |T_{\ell}(M_{i1},\cdots,M_{iq})-T_{\ell}(M'_{i1},\cdots,M'_{iq})|\leq \frac{1}{8}\sum_{i=1}^n\sum_{\ell,r=1}^q |M_{ir}-M'_{ir}| \nonumber\\
  &\leq& \frac{1}{8}\sum_{i,j=1}^n\sum_{\ell,r=1}^q |B_{ij}||1\{x_j=r\}-1\{x_j'=r\}|\nonumber\\
  &\leq& \frac{q\|B\|_1}{8}\sum_{j=1}^n\sum_{r=1}^q |1\{x_j=r\}-1\{x_j'=r\}|=\frac{q\|B\|_1}{4} H(\mathbf{x},\mathbf{x}'). 
\end{eqnarray}

By the path coupling lemma \cite[Corollary 14.8]{MR3726904}, for any $t\in\mathbb{N}^{*}$,
\begin{equation*}
    \max_{x\in [q]^n}\|K_x^t-\pi\|_{\mathrm{TV}}\leq n\Big(\frac{q\|B\|_1}{4}\Big)^t, 
\end{equation*}
%
and also for any $\epsilon\in (0,1)$,
\begin{equation*}
    t_{\mathrm{mix}}(\epsilon)\leq \Big\lceil \frac{-\log{\epsilon}+\log{n}}{-\log\big(q\|B\|_1\slash 4\big)} \Big\rceil.
\end{equation*}
%


\subsection{Proof of Theorem \ref{Thm3.2}}\label{subsec:proof:Thm3.2}

Suppose the current states are $\mathbf{x}=(x_1,\cdots,x_n),\mathbf{x}'=(x_1',\cdots,x_n')\in [q]^n$ with $\mathbf{x}\neq\mathbf{x}'$. We use the same coupling as used in the proof of Theorem \ref{Thm3.1} and assume the notations there. Now for any $i\in [n]$ and $\ell\in [q]$, by (\ref{Eq5.4}), we have  
\begin{equation*}
    |M_{i\ell}-M'_{i\ell}|\leq \sum_{j=1}^n |B_{ij}|\leq \|B\|_1. 
\end{equation*}
Using the above display, along with (\ref{Eq5.8}) and (\ref{Eq5.9}), we have for any $i\in [n]$ and $\ell\in [q]$, 
\begin{eqnarray*}
    \mathbb{P}(X_{i}'(1)=\ell|\mathbf{Z},\mathbf{Z}')&=&\frac{\exp(M'_{i\ell})}{\sum_{\ell'=1}^q \exp(M'_{i\ell'})}\geq \frac{\exp(M_{i\ell}-\|B\|_1)}{\sum_{\ell'=1}^q\exp(M_{i \ell'}+\|B\|_1)}\nonumber\\
    &=& \exp(-2\|B\|_1)\mathbb{P}(X_{i}(1)=\ell|\mathbf{Z},\mathbf{Z}').
\end{eqnarray*}
By (\ref{Eq5.1}) and the above display, for any $i\in [n]$,
\begin{eqnarray*}
    \mathbb{P}(X_{i}(1)= X_{i}'(1)|\mathbf{Z},\mathbf{Z}')&=&\sum_{\ell=1}^q \min\{\mathbb{P}(X_{i}(1)=\ell|\mathbf{Z},\mathbf{Z}'),\mathbb{P}(X'_{i}(1)=\ell|\mathbf{Z},\mathbf{Z}')\} \nonumber\\
    &\geq& \exp(-2\|B\|_1)\sum_{\ell=1}^q \mathbb{P}(X_{i}(1)=\ell|\mathbf{Z},\mathbf{Z}')=\exp(-2\|B\|_1),
\end{eqnarray*}
which gives
\begin{equation*}
    \mathbb{P}(X(1)=X'(1))=\mathbb{E}\Big[\prod_{i=1}^n\mathbb{P}(X_{i}(1)=X_{i}'(1)|\mathbf{Z},\mathbf{Z}')\Big]\geq \exp(-2n\|B\|_1). 
\end{equation*}

By the coupling lemma \cite[Theorem 5.4]{MR3726904}, for any $t\in\mathbb{N}^{*}$, we have
\begin{equation*}
 \max_{x\in [q]^n}\|K_x^t-\pi\|_{\mathrm{TV}}\leq (1-\exp(-2n\|B\|_1))^t\leq \exp(-t\exp(-2n\|B\|_1)).
\end{equation*}
Consequently, for any $\epsilon\in (0,1)$, $ t_{\mathrm{mix}}(\epsilon)\leq \lceil \exp(2\|B\|_1 n) \log(\epsilon^{-1})\rceil$.

\section{Proof of Theorem \ref{Thm3.3}}\label{sec:proof2}

In this section, we present the proof of Theorem \ref{Thm3.3}. We assume $\beta>q$ throughout this section. For any $\mathbf{x}=(x_1,\cdots,x_n)\in [q]^n$, we define $C(\mathbf{x}):=(C_1(\mathbf{x}),\cdots,C_q(\mathbf{x}))$, where for each $\ell\in [q]$, $C_{\ell}(\mathbf{x}):=n^{-1}\sum_{i=1}^n \mathbbm{1}_{x_i=\ell}$ is the proportion of color $\ell$ in $\mathbf{x}$.

Our proof of Theorem \ref{Thm3.3} consists of four main steps. Firstly, we show the convergence of a deterministic iteration in Section \ref{subsec:proof:Thm3.3_part2}. Then we show in Proposition \ref{P2.2} that within $\mathcal{O}(\log{n})$ iterations, there is a color whose proportion is greater than the proportion of any other color by a constant amount. Based on this, we show in Proposition \ref{P2.3} that the state evolution of the AG sampler is closely approximated by this deterministic iteration and, as a result, the macroscopic proportions of the $q$ colors are close to the expected proportions. Finally, after the macroscopic proportions of all colors have converged, we develop a novel coupling argument to show that the AG sampler mixes in $\mathcal{O}(\log{n})$ iterations in Section \ref{subsec:proof:Thm3.3_part4}.

\subsection{Notations and preliminary lemmas}\label{subsec:proof:Thm3.3_part1}

In this subsection, we present some notations and preliminary lemmas that will be used in later parts of the proof. We start with the following two definitions.

\begin{definition}\label{Def1.1}
We define $\mathcal{X}:=\big\{\boldsymbol \alpha = (\alpha_1,\cdots,\alpha_q):\alpha_1,\cdots,\alpha_q\geq 0, \sum_{\ell=1}^q \alpha_\ell=1\big\}$.
For any $\varepsilon\geq 0$, we define $\mathcal{X}_{\varepsilon}:=\{\bm{\alpha}=(\alpha_1,\cdots,\alpha_q)\in\mathcal{X}:\alpha_1\geq \cdots\geq \alpha_q, \alpha_1-\alpha_2\geq\varepsilon \}$.
For any $\bm{\alpha}=(\alpha_1,\cdots,\alpha_q)\in \mathbb{R}^q$, we define $\Phi(\bm{\alpha}):=(\Phi_1(\bm{\alpha}),\Phi_2(\bm{\alpha}),\cdots,\Phi_q(\bm{\alpha}))$ with 
\begin{equation*}
    \Phi_{\ell}(\bm{\alpha}):=\frac{e^{\beta \alpha_{\ell}}}{\sum_{s=1}^q e^{\beta \alpha_{s}}}=T_{\ell}(\beta\alpha_1,\beta\alpha_2,\cdots,\beta\alpha_q), \text{ for every } \ell\in [q],
\end{equation*}
where $T_{\ell}$ is as in \eqref{deftl}.
\end{definition}

Lemmas \ref{L2.1}-\ref{L2.4} below establish some basic properties of $\Phi(\cdot)$. The proofs of these lemmas are deferred to Appendix \ref{secB}.  


\begin{lemma}\label{L2.1}
Assume that $\beta>q$. Then the following conclusions hold:
\begin{itemize}
    \item[(a)] The equation
    \begin{align}\label{Eq2.3}
        e^{\beta(x-y)}=\frac{x}{y},\qquad x+(q-1)y=1, \qquad 0<y<x<1
\end{align}
has a unique solution $(a,b)$, say. Further, $(a,b)$  satisfy $a>q^{-1}$ and $\beta q a b<1$;
    \item[(b)] The function $g(x)=\frac{e^{\beta x}-1}{e^{\beta x}+q-1}$ is monotone increasing on $[0,1]$, and satisfies $g(0)=0$, $g(a-b)=a-b$, $g(x)>x$ when $x\in (0,a-b)$, and $g(x)<x$ when $x\in (a-b,1]$;
    \item[(c)] The function $h(x):=\frac{e^{\beta x}}{e^{\beta x}+(q-1)e^{\beta(1-x)\slash (q-1)}}$ is monotone increasing on $[q^{-1},1]$, and satisfies $h(q^{-1})=q^{-1}$, $h(a)=a$, $h(x)>x$ when $x\in (q^{-1},a)$, and $h(x)<x$ when $x\in (a,1]$.
\end{itemize}
\end{lemma}

\begin{definition}\label{Def1.2}
We define $\bm{\alpha}_0:=(a,b,\cdots,b)\in\mathcal{X}_0$, where $a,b$ are as in Lemma~\ref{L2.1}(a). Note that $\Phi(\bm{\alpha}_0)=\bm{\alpha}_0$.
\end{definition}
\begin{lemma}\label{L1.2}
For any $\bm{\alpha},\bm{\alpha}'\in\mathbb{R}^q$, we have $\|\Phi(\bm{\alpha}')-\Phi(\bm{\alpha})\|_1\leq   2\beta q\|\bm{\alpha'}-\bm{\alpha}\|_{\infty}$.
\end{lemma}


\begin{lemma}\label{L1.4}
There exist positive constants $\rho_0,\delta_0\in (0,1)$ that only depend on $q,\beta$, such that for any $\bm{\alpha}=(\alpha_1,\cdots,\alpha_q),\bm{\alpha}'=(\alpha'_1,\cdots,\alpha'_q)\in \mathbb{R}^q$ with $\max\{\|\bm{\alpha}-\bm{\alpha}_0\|_2,\|\bm{\alpha}'-\bm{\alpha}_0\|_2\}\leq \delta_0$ and $\sum_{\ell=1}^q\alpha_{\ell}=\sum_{\ell=1}^q \alpha_{\ell}'$, we have $\|\Phi(\bm{\alpha}')-\Phi(\bm{\alpha})\|_1\leq \rho_0 \|\bm{\alpha}'-\bm{\alpha}\|_1$.
\end{lemma}




\begin{lemma}\label{L2.4}
For any $\rho\in (0,1)$, there exists a constant $\kappa_{\rho}\in (0,1)$ that only depends on $q,\beta,\rho$, such that for any $\bm{\alpha}=(\alpha_1,\cdots,\alpha_q)\in\mathbb{R}^d$ such that $\alpha_1\geq \alpha_2\geq\cdots\geq\alpha_q$, we have $\Phi_1(\bm{\alpha})-\Phi_2(\bm{\alpha})  \geq (\rho \beta\slash q)\min\{\alpha_1-\alpha_2,\kappa_{\rho}\}$.
\end{lemma}


\subsection{Convergence of a deterministic iteration}\label{subsec:proof:Thm3.3_part2}

In this subsection, we introduce the following convenient notation. For any function \mbox{$f: \mathcal X \to \mathcal X$}, we denote by $f^{(t)}$ the function obtained by applying $t$ times the function $f$, that is $f^{(t)} (x) = f \circ f \circ \cdots \circ f(x)$. 
For example, $\Phi^{(t)}$ is the function obtained by applying $t$ times $\Phi: \mathbb R^q \to \mathbb R^q$.
Furthermore, we write $\boldsymbol \alpha(t) = \Phi^{(t)}(\boldsymbol \alpha)$.
We will now show uniform convergence of the deterministic iteration $\{\Phi^{(t)}(\bm{\alpha})\}_{t=1}^{\infty}$ for $\bm{\alpha}\in\mathcal{X}_{\varepsilon}$ for any $\varepsilon\in (0,a-b)$.
The main result is given by the following proposition.




\begin{proposition}\label{P2.1}
For any $\varepsilon\in (0,a-b)$, we have $\lim\limits_{t\rightarrow\infty} \sup\limits_{\bm{\alpha}\in\mathcal{X}_{\varepsilon}}\|\Phi^{(t)}(\bm{\alpha})-\pmb{\alpha}_0\|_2=0$.
\end{proposition}

The rest of this subsection is devoted to the proof of Proposition \ref{P2.1}. Recall $g$ and $h$ from Lemma~\ref{L2.1}(b)-(c). 
For any $t \in \mathbb  N$, we have
\begin{equation*}
   \alpha_{1}(t+1)-\alpha_{2}(t+1)=\frac{e^{\beta\alpha_{1}(t)}-e^{\beta\alpha_{2}(t)}}{\sum_{\ell=1}^q e^{\beta\alpha_{\ell}(t)}}\geq \frac{e^{\beta\alpha_{1}(t)}-e^{\beta\alpha_{2}(t)}}{e^{\beta\alpha_{1}(t)}+(q-1)e^{\beta\alpha_{2}(t)}}=g(\alpha_{1}(t)-\alpha_{2}(t)).
\end{equation*}
By Jensen's inequality,
\begin{equation*}
    \frac{1}{q-1}\sum_{\ell = 2}^q e^{\beta \alpha_{\ell}(t)} \geq e^{\beta\big(\sum_{\ell=2}^q \alpha_{k}(t)\big)\big\slash(q-1)}=e^{\beta(1-\alpha_{1}(t))\slash (q-1)}.
\end{equation*}
Hence
\begin{equation*}
    \alpha_{1}(t+1)=\frac{e^{\beta \alpha_{1}(t)}}{\sum_{\ell=1}^q e^{\beta \alpha_{\ell}(t)}}\leq \frac{e^{\beta \alpha_{1}(t)}}{e^{\beta \alpha_{1}(t)}+(q-1)e^{\beta(1-\alpha_{1}(t))\slash (q-1)}}=h(\alpha_{1}(t)).
\end{equation*}
As $g$ is monotone increasing on $[0,1]$ and $h$ is monotone increasing on $[q^{-1},1]$ (by Lemma \ref{L2.1}), by inductively applying the above displays, for any $t\in \mathbb{N}$, we get
\begin{equation}\label{Eq2.7}
    \alpha_{1}(t)-\alpha_{2}(t)\geq g^{(t)}(\alpha_1-\alpha_2)\geq g^{(t)}(\varepsilon), \quad \alpha_{1}(t)\leq h^{(t)}(\alpha_1)\leq h^{(t)}(1).
\end{equation}

Again using the monotonicity of $g$ from Lemma \ref{L2.1}, for any $\varepsilon\in (0,a-b)$ we get
$$0=g(0)\le g(\varepsilon)\le g(a-b)=a-b.$$
By induction, we have $g^{(t)}(\varepsilon)\in(0, a-b)$ for all $t \in \mathbb{N}$. Furthermore, $t\mapsto g^{(t)}(\varepsilon)$ is non-decreasing in $t$, as $g(x)>x$ for all $x\in (0,a-b)$ (see Lemma \ref{L2.1}). Consequently, the sequence $g^{(t)}(\varepsilon)$ has a limit $g^{(\infty)}(\varepsilon),$ say. 

In a similar manner, one can show that $h^{(t)}(1)\ge a$, and is non-increasing in $t$. Hence  $\lim_{t\rightarrow\infty}h^{(t)}(1)$ exists, and equals $h^{(\infty)}(1)$, say. Taking $t\rightarrow\infty$ on both sides of $g(g^{(t)}(\varepsilon))=g^{(t+1)}(\varepsilon)$ and $h(h^{(t)}(1))=h^{(t+1)}(1)$, we obtain that
\begin{equation*}
    g(g^{(\infty)}(\varepsilon))=g^{(\infty)}(\varepsilon), \quad h(h^{(\infty)}(1))=  h^{(\infty)}(1).
\end{equation*}
Thus $g^{(\infty)}(\varepsilon)$ and $h^{(\infty)}(1)$ are fixed points of $g(\cdot)$ and $h(\cdot)$ respectively, and using Lemma \ref{L2.1} gives $g^{(\infty)}(\varepsilon)\in \{0, a-b\}$, and $h^{(\infty)}(1)\in \{q^{-1},a\}$. Finally, 
using the fact that $t\mapsto g^{(t)}(\varepsilon)$ is monotone, and $h^{(t)}(1)\ge a$ (proved above), we get
\begin{equation*}
   g^{(\infty)}(\varepsilon)=\lim_{t\rightarrow\infty} g^{(t)}(\varepsilon)\geq g^{(0)}(\varepsilon)=\varepsilon>0,\qquad h^{(\infty)}(1)=\lim_{t\rightarrow\infty}h^{(t)}(1)\ge a>q^{-1}.
\end{equation*}
Hence we have
$$g^{(\infty)}(\varepsilon)=a-b,\qquad  h^{(\infty)}(1)=a.$$

Using the above display along with (\ref{Eq2.7}), for any $\delta>0$ there exists $T_{\delta,\varepsilon}\in\mathbb{N}^{*}$ that only depends on $\delta,\varepsilon$, such that for any $t\geq T_{\delta,\varepsilon}$ and $\bm{\alpha}=(\alpha_1,\cdots,\alpha_q)\in \mathcal{X}_{\varepsilon}$, 
\begin{equation*}\label{Eq2.13}
    \alpha_{1}(t)-\alpha_{2}(t)\geq a-b-\delta\slash (2q), \quad \alpha_{1}(t)\leq a+\delta\slash (2q).
\end{equation*}
The above display implies $\alpha_{2}(t)\leq b+\delta\slash q$, and consequently,
\begin{align*}
    \alpha_{\ell}(t)\leq \alpha_{2}(t)\leq b+\delta\slash q, \quad\text{ for any }  \ell\in [2,q]\cap\mathbb{N}.
\end{align*}
Using the above two displays, we get the following lower bounds:
\begin{eqnarray*}\label{Eq2.14}
   \alpha_{1}(t)&=&1-\sum_{\ell=2}^q \alpha_{\ell}(t)\geq 1-\sum_{\ell=2}^q(b+\delta\slash q)=1-(q-1)b-\frac{q-1}{q}\delta\geq a-\delta,\nonumber\\
    \alpha_{\ell}(t)&=&1-\sum_{\ell'\in [q]:\ell'\neq \ell}\alpha_{\ell'}(t)\geq 1-(a+\delta\slash (2q))-(q-2)(b+\delta\slash q)\geq b-\delta,
\end{eqnarray*}
for any $\ell\in [2,q]\cap\mathbb{N}$. Using the last four displays, for any $t\geq T_{\delta,\varepsilon}$ and $\bm{\alpha}=(\alpha_1,\cdots,\alpha_q)\in \mathcal{X}_{\varepsilon}$,
\begin{equation*}
    \|\Phi^{(t)}(\bm{\alpha})-(a,b,\cdots,b)\|_2\leq \sqrt{q}\|\Phi^t(\bm{\alpha})-(a,b,\cdots,b)\|_{\infty}\leq \sqrt{q}\delta.
\end{equation*}
Hence $\lim\limits_{t\rightarrow\infty} \sup\limits_{\bm{\alpha}\in\mathcal{X}_{\varepsilon}}\|\Phi^{(t)}(\bm{\alpha})-(a,b,\cdots,b)\|_2=0$.

\subsection{Convergence to the expected color proportions}\label{subsec:proof:Thm3.3_part3}


In this subsection, we state and prove two propositions, which are the major technical steps in proving Theorem \ref{Thm3.3}.

Throughout the remainder of this section, we denote by $C$ and $c$ positive constants that depend only on $q,\beta$. The values of these constants may change from line to line.

The iterations of the modified AG sampler for the Curie-Weiss model can be described as follows. For each $t\in\mathbb{N}$, we denote by $\mathbf{X}(t)=(X_1(t),\cdots,X_n(t))\in [q]^n$ the state at step $t$. Given $\mathbf{X}(t)$, we set $\mathbf{Y}(t):=(Y_{\ell}(t))_{\ell\in [q]}$ with $Y_{\ell}(t)=n^{-1}\sum_{i=1}^n 1\{X_i(t)=\ell\}$, and perform the following steps to generate $\mathbf{X}(t+1)$:
\begin{itemize}
    \item For each $\ell\in[q]$, we independently sample $W_{\ell}(t)\sim N(0,(n\beta)^{-1})$ and take $Z_{\ell}(t)=Y_{\ell}(t)+W_{\ell}(t)$ (note that $Z_{\ell}(t)\sim N(Y_{\ell}(t),(n\beta)^{-1})$). We set $\mathbf{Z}(t):=(Z_{\ell}(t))_{\ell\in[q]}$.
    \item For each $i\in [n]$, we independently sample $\widetilde{X}_i(t+1)\in [q]$ such that for each $\ell \in  [q]$,
    \begin{equation*}
         \mathbb{P}(\widetilde{X}_{i}(t+1)=\ell\mid \mathbf{Z}(t))=\frac{e^{\beta  Z_{\ell}(t)}}{\sum_{\ell'=1}^q e^{\beta Z_{\ell'}(t)}}.
    \end{equation*}
    Then we sample a uniform random permutation $\tau_t\in S_q$, and set $\mathbf{X}(t+1)=(X_1(t+1),\cdots,X_n(t+1))$ with $X_{i}(t+1)=\tau_t(\widetilde{X}_i(t+1))$.
\end{itemize}





For any $t\in \mathbb{N}$, we denote by $Y_{(1)}(t)\geq Y_{(2)}(t)\geq\cdots\geq Y_{(q)}(t)$ the order statistics of $Y_{1}(t),Y_{2}(t),\cdots,Y_{q}(t)$ and $Z_{(1)}(t)\geq Z_{(2)}(t)\geq\cdots \geq  Z_{(q)}(t)$ the order statistics of $Z_{1}(t),Z_{2}(t),\cdots,Z_{q}(t)$. For any $t\in \mathbb{N}^{*}$ and $\ell\in [q]$, we let $\widetilde{Y}_{\ell}(t):=n^{-1}\sum_{i=1}^n 1\{\widetilde{X}_{i}(t)=\ell\}$, and note that
\begin{equation}\label{Eq3.4}
    \widetilde{Y}_{\ell}(t)=\frac{1}{n}\sum_{i=1}^n1\{X_{i}(t) =\tau_t(\ell)\}=Y_{\tau_t(\ell)}(t).
\end{equation}

\begin{definition}[The filtrations $\mathcal{F}_t,\mathcal{G}_t$ and the events $\mathcal{E}_{t,u,v},\mathcal{A}_{t,u,v}$]\label{def_EA}
For any $t\in\mathbb{N}$, let $\mathcal{F}_t$ be the $\sigma$-algebra generated by $\mathbf{X}(0),\mathbf{Z}(0),\cdots,\mathbf{X}(t-1),\mathbf{Z}(t-1),\mathbf{X}(t)$, 
and let $\mathcal{G}_t$ be the $\sigma$-algebra generated by $\mathbf{X}(0),\mathbf{Z}(0),\cdots,\mathbf{X}(t), \mathbf{Z}(t)$. 

For any $t\in \mathbb{N}$ and $u,v\geq 0$, let $\mathcal{E}_{t,u,v}$ 
be the event that for every $\ell\in [q]$, we have
\begin{equation*}
    |Z_{\ell}(t)-Y_{\ell}(t)|\leq u, \qquad \bigg|\widetilde{Y}_{\ell}(t+1)-\frac{e^{\beta Z_{\ell}(t)}}{\sum_{\ell'=1}^q e^{\beta Z_{\ell'}(t)}}\bigg|\leq v.
\end{equation*}
Let $\kappa:=\kappa_{\sqrt{q\slash \beta}}$ be defined as in Lemma \ref{L2.4}. Let $\mathcal{A}_{t,u,v}$ be the event that both of the following hold:
\begin{equation*}
    Y_{(1)}(t+1)-Y_{(2)}(t+1)\geq\sqrt{\beta\slash q}\min\{Y_{(1)}(t)-Y_{(2)}(t),\kappa\}-2\sqrt{\beta}(u+v),
\end{equation*}
\begin{equation*}
    Y_{(1)}(t+1)-Y_{(2)}(t+1)\geq\sqrt{\beta\slash q}\min\{Z_{(1)}(t)-Z_{(2)}(t),\kappa\}-2\sqrt{\beta}v.
\end{equation*}
\end{definition}

Based on Lemma \ref{L2.4}, we establish the following two lemmas. 

\begin{lemma}\label{L2.6}
For any $t\in \mathbb{N}$, $\ell\in [q]$, and $u,v\geq 0$,
\begin{align*}
    \mathbb{P}(|Z_{\ell}(t)-Y_{\ell}(t)|\geq& u|\mathcal{F}_t)\leq 2\exp(-n\beta u^2\slash 2),\\
    \mathbb{P}\Big(\Big|\widetilde{Y}_{\ell}(t+1)-\frac{e^{\beta Z_{\ell}(t)}}{\sum_{\ell'=1}^q e^{\beta Z_{\ell'}(t)}}\Big|\geq& v\Big|\mathcal{G}_t\Big)\leq 2\exp(-2n v^2).
\end{align*}
Consequently, we have
\begin{equation*}
     \mathbb{P}((\mathcal{E}_{t,u,v})^c|\mathcal{F}_t)\leq 2q\exp(-n\beta u^2\slash 2)+2q\exp(-2n v^2).
\end{equation*}
\end{lemma}
\begin{proof}
The two inequalities follow by tail bounds for Gaussian random variables and Hoeffding's inequality, respectively.
\end{proof}

\begin{lemma}\label{L2.5}
For any $t\in \mathbb{N}$ and $u,v\geq 0$, we have $\mathcal{E}_{t,u,v}\subseteq\mathcal{A}_{t,u,v}$. 
\end{lemma}

\begin{proof}
    By (\ref{Eq3.4}) and the rearrangement inequality, on the set $\mathcal{E}_{t,u,v}$ we have
\begin{align*}
    \sum_{\ell=1}^q \bigg(Y_{(\ell)}(t+1)-\frac{e^{\beta Z_{(\ell)}(t)}}{\sum_{\ell'=1}^q e^{\beta Z_{\ell'}(t)}}\bigg)^2\leq & \sum_{\ell=1}^q \bigg(\widetilde{Y}_{\ell}(t+1)-\frac{e^{\beta Z_{\ell}(t)}}{\sum_{\ell'=1}^q e^{\beta Z_{\ell'}(t)}}\bigg)^2\leq q v^2.
\end{align*}
Using the triangle inequality along with the above bound gives 
\begin{eqnarray*}
  &&  Y_{(1)}(t+1)-Y_{(2)}(t+1)\nonumber\\
  &\geq& \frac{e^{\beta Z_{(1)}(t)}-e^{\beta Z_{(2)}(t)}}{\sum_{\ell'=1}^q e^{\beta Z_{\ell'}(t)}}-\bigg|Y_{(1)}(t+1)-\frac{e^{\beta Z_{(1)}(t)}}{\sum_{\ell'=1}^q e^{\beta Z_{\ell'}(t)}}\bigg|-\bigg|Y_{(2)}(t+1)-\frac{e^{\beta Z_{(2)}(t)}}{\sum_{\ell'=1}^q e^{\beta Z_{\ell'}(t)}}\bigg|\nonumber\\
  &\geq&\frac{e^{\beta Z_{(1)}(t)}-e^{\beta Z_{(2)}(t)}}{\sum_{\ell'=1}^q e^{\beta Z_{\ell'}(t)}}-2\sqrt{q}v\geq\sqrt{\beta\slash q}\min\{Z_{(1)}(t)-Z_{(2)}(t),\kappa\}-2\sqrt{q}v,
\end{eqnarray*}
where the last inequality uses Lemma \ref{L2.4} with $\rho=\sqrt{q\slash \beta}$. The RHS above can be bounded below by
\begin{equation*}
  \sqrt{\beta\slash q}\min\{Y_{(1)}(t)-Y_{(2)}(t)-2\sqrt{q}u,\kappa\}-2\sqrt{q}v \geq \sqrt{\beta\slash q}\min\{Y_{(1)}(t)-Y_{(2)}(t),\kappa\}-2\sqrt{\beta}(u+v), 
\end{equation*}
where in the first bound we again use the rearrangement inequality to conclude that
\begin{equation*}
    \sum_{\ell=1}^q (Z_{(\ell)}(t)-Y_{(\ell)}(t))^2\leq \sum_{\ell=1}^q (Z_{\ell}(t)-Y_{\ell}(t))^2\leq qu^2.
\end{equation*}
\end{proof}

\begin{proposition}\label{P2.2}
There exist positive constants $C_1,C_2,c_1,c_2$ that only depend on $q,\beta$, such that the following holds. Let $\mathcal{H}_1$ be the event that $Y_{(1)}(t)-Y_{(2)}(t)\geq c_1$ for some $t\in [1,C_1\log(n+1)]\cap\mathbb{N}$. Then for any random starting state $\mathbf{X}(0)\in [q]^n$, we have
\begin{equation*}
    \mathbb{P}(\mathcal{H}_1^c)\leq C_2 \exp(-c_2 \log\log(n+1)).
\end{equation*}
\end{proposition}
\begin{proof}
Without loss of generality, we assume that $n\geq 20$. The proof is divided into the following two steps.

\paragraph{Step 1}


For any $t\in \mathbb{N}$, let $I_{t}\in S_q$ be such that $Y_{I_{t}(1)}(t)\geq Y_{I_{t}(2)}(t)\geq \cdots \geq Y_{I_{t}(q)}(t)$. Note that for any $t\in\mathbb{N}^{*}$ and $x\geq 0$, if $W_{I_{t}(1)}(t)-W_{I_{t}(2)}(t)\geq x$ and $\max_{\ell\in [q]\backslash [2]}  W_{I_{t}(\ell)}(t)\leq W_{I_{t}(2)}(t)$, then 
\begin{align*}
Z_{I_t(1)}(t)-x =Y_{I_t(1)}(t)+W_{I_t(1)}(t)-x&\geq Y_{I_t(2)}(t)+W_{I_t(2)}(t)\nonumber\\
& \geq \max_{\ell\in[q]\backslash [2]}\big\{Y_{I_t(\ell)}(t)+W_{I_t(\ell)}(t)\big\}.
\end{align*}
In particular, this implies $Z_{(\ell)}(t)=Z_{I_t(\ell)}(t)$ for $\ell=1,2$, and so $Z_{(1)}(t)-Z_{(2)}(t)=Z_{I_t(1)}(t)-Z_{I_t(2)}(t)\geq x$. Consequently,
\begin{equation*}
   \Big\{W_{I_{t}(1)}(t)-W_{I_{t}(2)}(t)\geq x, \max_{\ell\in [q]\backslash [2]}  W_{I_{t}(\ell)}(t)\leq W_{I_{t}(2)}(t)  \Big\} \subseteq \{Z_{(1)}(t)-Z_{(2)}(t)\geq x\},
\end{equation*}
which in turn implies
\begin{eqnarray*}\label{Eq3.1}
  && \mathbb{P}(Z_{(1)}(t)-Z_{(2)}(t)\geq x|\mathcal{F}_{t})\geq \mathbb{P}(W_1-W_2\geq x, \max_{\ell\in [q]\backslash [2]}W_{\ell}\leq W_2) \nonumber\\
  &=& \int_{-\infty}^{\infty} \phi(y)(1-\Phi(y+\sqrt{n\beta}x))\Phi(y)^{q-2} dy\geq \int_{-1}^0 \phi(y)(1-\Phi(y+\sqrt{n\beta}x))\Phi(y)^{q-2} dy,
\end{eqnarray*}
where $\sqrt{n\beta}\{W_{\ell}\}_{\ell\in [q]}$ are i.i.d. $N(0,1)$ random variables, and $\phi(\cdot),\Phi(\cdot)$ are respectively the probability density function and cumulative distribution function of $N(0,1)$.
By \cite{MR5558}, for any $s\geq 1$, 
\begin{equation*}\label{Eq3.2}
    1-\Phi(s)\geq \frac{s}{\sqrt{2\pi}(s^2+1)}e^{-s^2\slash 2}\geq cs^{-1}e^{-s^2\slash 2}.
\end{equation*}
Note that for any $y\in[-1,0]$, $1-\Phi(y+\sqrt{n\beta}x)\geq 1-\Phi(\sqrt{n\beta}x)$ and $\Phi(y)\geq \Phi(-1) \geq c$. Hence using the above two displays, for any $t\in\mathbb{N}^{*}$ and $x\geq (n\beta)^{-1\slash 2}$, we get 
\begin{eqnarray*}
    \mathbb{P}(Z_{(1)}(t)-Z_{(2)}(t)\geq x|\mathcal{F}_{t})&\geq& c(1-\Phi(\sqrt{n\beta}x))\int_{-1}^0\phi(y)dy\geq c(1-\Phi(\sqrt{n\beta}x))\nonumber\\
   &\geq& c(n\beta)^{-1\slash 2}x^{-1}\exp(-n\beta x^2\slash 2)\geq c\exp(-n\beta x^2).
\end{eqnarray*}
Hence for any $x\geq (n\beta)^{-1\slash 2}$,
\begin{eqnarray}\label{Eq3.3}
    \mathbb{P}\Big(\max_{t\in [1,2\log{n}]\cap\mathbb{N}}\{Z_{(1)}(t)-Z_{(2)}(t)\}< x\Big)&\leq& (1-c\exp(-n\beta x^2))^{\lfloor 2\log{n} \rfloor}\nonumber\\
    &\leq& \exp(-c\log{n}\exp(-n\beta x^2)).
\end{eqnarray}

Let
\begin{equation}\label{def_T1}
    T_1:=\inf\Bigg\{t\in\mathbb{N}^{*}:Z_{(1)}(t)-Z_{(2)}(t)\geq \frac{1}{2}\sqrt{\frac{\log\log{n}}{n\beta}}\Bigg\}.
\end{equation}
Let $\mathcal{D}_1$ be the event that $T_1\leq 2\log{n}$. By (\ref{Eq3.3}), 
\begin{eqnarray}\label{Eq3.18}
    \mathbb{P}(\mathcal{D}_1^c)&\leq& \mathbb{P}\left(\max_{t\in [1,2\log{n}]\cap\mathbb{N}}\{Z_{(1)}(t)-Z_{(2)}(t)\}< \frac{1}{2}\sqrt{\frac{\log\log{n}}{n\beta}}\right)\nonumber\\
  &\leq& \exp(-c\log{n}\exp(-\log\log{n}\slash 4))\leq \exp(-c\sqrt{\log{n}}).
\end{eqnarray}



\paragraph{Step 2}

Consider any $t\in \mathbb{N}$. For any $u,v\geq 0$, let $\kappa,\mathcal{E}_{t,u,v},\mathcal{A}_{t,u,v}$ be as in Definition \ref{def_EA}. 
By Lemma \ref{L2.5}, we have $\mathcal{E}_{t,u,v}\subseteq\mathcal{A}_{t,u,v}$, and so 
\begin{equation*}\label{Eq3.8}
     \mathbb{P}((\mathcal{A}_{t,u,v})^c|\mathcal{F}_t)\leq\mathbb{P}((\mathcal{E}_{t,u,v})^c|\mathcal{F}_t)
    \leq 2q\Big[\exp(-nu^2)+\exp(-nv^2)\Big]\leq 4q\exp(-n\min\{u,v\}^2),
\end{equation*}
where the second inequality uses Lemma \ref{L2.6}.  
For any $u\geq 0$, let $\mathcal{B}_{t,u}$ be the event that $T_1<\infty$ and
\begin{equation*}
    Y_{(1)}(T_1+t+1)-Y_{(2)}(T_1+t+1)\geq \sqrt{\beta\slash q} \min\{Y_{(1)}(T_1+t)-Y_{(2)}(T_1+t),\kappa\}-4\sqrt{\beta}u,
\end{equation*}
\begin{equation*}
    Y_{(1)}(T_1+t+1)-Y_{(2)}(T_1+t+1)\geq \sqrt{\beta\slash q}\min\{Z_{(1)}(T_1+t)-Z_{(2)}(T_1+t),\kappa\}-2\sqrt{\beta}u.
\end{equation*}
By the strong Markov property (see, for example, \cite[Proposition 6.1.16]{dembo2021probability}), we have
\begin{equation*}
    \mathbb{P}((\mathcal{B}_{t,u})^c|\mathcal{F}_{T_1+t})\mathbbm{1}_{T_1<\infty}=\mathbb{P}((\mathcal{A}_{t,u,u})^c|\mathcal{F}_t)\mathbbm{1}_{T_1<\infty}\leq 4q\exp(-n u^2),
\end{equation*}
hence
\begin{equation}\label{Eq3.15}
     \mathbb{P}((\mathcal{B}_{t,u})^c\cap\{T_1<\infty\})\leq 4q\exp(-nu^2).
\end{equation}

Let
\begin{equation}\label{def_L}
    L:=\min\bigg\{t\in\mathbb{N}: \frac{1}{2}\Big(\frac{\beta}{q}\Big)^{t\slash 4}\sqrt{\frac{\log\log{n}}{n\beta}}\geq\kappa\bigg\}.
\end{equation}
For any $t\in\mathbb{N}$, we let 
\begin{equation}
    u_t:=\frac{1}{4\sqrt{\beta}}\Big(\Big(\frac{\beta}{q}\Big)^{1\slash 2}-\Big(\frac{\beta}{q}\Big)^{1\slash 4}\Big)\Big(\frac{\beta}{q}\Big)^{t\slash 4}\min\bigg\{\frac{1}{2}\sqrt{\frac{\log\log{n}}{n\beta}},\kappa\bigg\}.
\end{equation}
We also let $\mathcal{D}_2:=\bigcap_{t\in [0,L]\cap\mathbb{N}}\mathcal{B}_{t,u_t}$. By (\ref{Eq3.15}) and the union bound,
\begin{eqnarray}\label{Eq3.16}
   \mathbb{P}(\mathcal{D}_2^c\cap\{T_1<\infty\})&\leq& \sum_{t\in [0,L]\cap\mathbb{N}}\mathbb{P}((\mathcal{B}_{t,u_t})^c\cap\{T_1<\infty\}) \leq 4q\sum_{t\in [0,L]\cap\mathbb{N}}\exp(-nu_t^2)\nonumber\\
   &\leq& C\sum_{t=0}^{\infty}\exp(-c(\beta\slash q)^{t\slash 2}\log\log{n})\leq C\exp(-c\log\log{n}).
\end{eqnarray}

Assume that the event $\mathcal{D}_1\cap\mathcal{D}_2$ holds. By the definition of $T_1$ (see \eqref{def_T1}), we have $Z_{(1)}(T_1)-Z_{(2)}(T_1)\geq \frac{1}{2}\sqrt{\frac{\log\log{n}}{n\beta}}$. If $L=0$, we have $\frac{1}{2}\sqrt{\frac{\log\log{n}}{n\beta}}\geq\kappa$, hence on the event $\mathcal{B}_{0,u_0}\supseteq \mathcal{D}_2$, we have
\begin{eqnarray}\label{Eq3.11}
    &&Y_{(1)}(T_1+1)-Y_{(2)}(T_1+1)\geq \sqrt{\beta\slash q}\min\{Z_{(1)}(T_1)-Z_{(2)}(T_1),\kappa\}-2\sqrt{\beta}u_0 \nonumber\\
    &\geq&  \sqrt{\beta\slash q}\kappa-\frac{1}{2}\Big(\Big(\frac{\beta}{q}\Big)^{1\slash 2}-\Big(\frac{\beta}{q}\Big)^{1\slash 4}\Big)\kappa=\frac{\kappa}{2}\bigg(\Big(\frac{\beta}{q}\Big)^{1\slash 2}+\Big(\frac{\beta}{q}\Big)^{1\slash 4}\bigg)\geq\kappa.
\end{eqnarray}
If $L\geq 1$, we have (see the definition of $L$ in \eqref{def_L})
\begin{equation}\label{kappalowerbdd}
    \kappa>\frac{1}{2}\Big(\frac{\beta}{q}\Big)^{(L-1)\slash 4}\sqrt{\frac{\log\log{n}}{n\beta}}.
\end{equation}
Below, we show by induction that
\begin{equation}\label{Eq3.12}
    Y_{(1)}(T_1+t)-Y_{(2)}(T_1+t)   \geq\frac{1}{2}\Big(\frac{\beta}{q}\Big)^{t\slash 4}\sqrt{\frac{\log\log{n}}{n\beta}}
\end{equation}
for any $t\in [L]$ as follows. When $t=1$, on the event $\mathcal{B}_{0,u_0}\supseteq \mathcal{D}_2$, we have
\begin{eqnarray*}
    && Y_{(1)}(T_1+1)-Y_{(2)}(T_1+1) \geq \sqrt{\beta\slash q}\min\{Z_{(1)}(T_1)-Z_{(2)}(T_1),\kappa\}-2\sqrt{\beta}u_0 \nonumber\\
    &\geq& \frac{1}{2}\Big(\frac{\beta}{q}\Big)^{1\slash 2}\sqrt{\frac{\log\log{n}}{n\beta}}-\frac{1}{2}\Big(\Big(\frac{\beta}{q}\Big)^{1\slash 2}-\Big(\frac{\beta}{q}\Big)^{1\slash 4}\Big)\sqrt{\frac{\log\log{n}}{n\beta}}=\frac{1}{2}\Big(\frac{\beta}{q}\Big)^{1\slash 4}\sqrt{\frac{\log\log{n}}{n\beta}},
\end{eqnarray*}
where we use \eqref{kappalowerbdd} in the second inequality. This verifies the induction claim when $t=1$. Now suppose that (\ref{Eq3.12}) holds for some $t\in [L-1]$. On the event $\mathcal{B}_{t,u_t}\supseteq \mathcal{D}_2$, we have 
\begin{eqnarray*}
   && Y_{(1)}(T_1+t+1)-Y_{(2)}(T_1+t+1)\nonumber\\
   &\geq&\sqrt{\beta\slash q} \min\{Y_{(1)}(T_1+t)-Y_{(2)}(T_1+t),\kappa\}-4\sqrt{\beta}u_t \nonumber\\
   &\geq& \frac{1}{2}\Big(\frac{\beta}{q}\Big)^{1\slash 2}\Big(\frac{\beta}{q}\Big)^{t\slash 4}\sqrt{\frac{\log\log{n}}{n\beta}}-\frac{1}{2}\Big(\Big(\frac{\beta}{q}\Big)^{1\slash 2}-\Big(\frac{\beta}{q}\Big)^{1\slash 4}\Big)\Big(\frac{\beta}{q}\Big)^{t\slash 4}\sqrt{\frac{\log\log{n}}{n\beta}}\nonumber\\
   &=& \frac{1}{2}\Big(\frac{\beta}{q}\Big)^{(t+1)\slash 4}\sqrt{\frac{\log\log{n}}{n\beta}},
\end{eqnarray*}
where we use \eqref{kappalowerbdd} and \eqref{Eq3.12} in the second inequality. Hence by induction, \eqref{Eq3.12} holds for every $t\in [L]$, and consequently,
\begin{equation}\label{Eq3.13}
    Y_{(1)}(T_1+L)-Y_{(2)}(T_1+L)\geq \frac{1}{2}\Big(\frac{\beta}{q}\Big)^{L\slash 4}\sqrt{\frac{\log\log{n}}{n\beta}}\geq \kappa,
\end{equation}
where the last inequality follows from the definition of $L$ (see \eqref{def_L}). By (\ref{Eq3.11}) and (\ref{Eq3.13}), we get $Y_{(1)}(T_1+\max\{L,1\})-Y_{(2)}(T_1+\max\{L,1\})\geq \kappa$.
Moreover, note that on the event $\mathcal{D}_1$, 
\begin{equation*}
     T_1+\max\{L,1\}\leq 2\log{n}+\max\{L,1\}\leq C_1\log{n},
\end{equation*}
where $C_1$ is a positive constant that depends only on $q,\beta$. Taking $c_1=\kappa$ and recalling the definition of $\mathcal{H}_1$ from the statement of the proposition, we get
\begin{equation}\label{Eq3.17}
    \mathcal{D}_1\cap\mathcal{D}_2\subseteq \mathcal{H}_1 .
\end{equation}

Combining (\ref{Eq3.18}), (\ref{Eq3.16}), and (\ref{Eq3.17}), we conclude that
\begin{equation*}
    \mathbb{P}(\mathcal{H}_1^c)\leq \mathbb{P}(\mathcal{D}_1^c)+\mathbb{P}(\mathcal{D}_2^c\cap\mathcal{D}_1)\leq \mathbb{P}(\mathcal{D}_1^c)+\mathbb{P}(\mathcal{D}_2^c\cap\{T_1<\infty\})\leq C\exp(-c\log\log{n}).
\end{equation*}




\end{proof}

For any vector $\mathbf{x}=(x_1,\cdots,x_q)\in \mathbb{R}^q$, we let $s(\mathbf{x}):=(x_{(1)},\cdots,x_{(q)})$ be the order statistics of $\mathbf{x}$. 


\begin{proposition}\label{P2.3}
Let $\pmb{\alpha}_0$ be as in Definition \ref{Def1.2}, and let $C_1,c_1$ and the event $\mathcal{H}_1$ be defined as in Proposition \ref{P2.2}. For any $t\in\mathbb{N}$ and $\delta\in (0,1)$, let $\mathcal{V}_{t,\delta}$ be the event that $\|s(\mathbf{Y}(t))-\bm{\alpha}_0\|_1\leq\delta$. Then for any $K\geq 1$ and $\delta\in (0,1)$, there exist positive constants $C',c'$ that depend only on $q,\beta,K,\delta$, such that for any $t\in [(C_1+1)\log(n+1),K\log(n+1)]\cap\mathbb{N}$, we have
\begin{equation*}
    \mathbb{P}((\mathcal{V}_{t,\delta})^c\cap\mathcal{H}_1)\leq C'\exp(-c'n).
\end{equation*}
\end{proposition}
\begin{proof}
Let $\rho_0,\delta_0$ be defined as in Lemma \ref{L1.4}.
Without loss of generality, we assume that $\delta\leq \delta_0$ and $c_1<a-b$. Let $T_2:=\inf\{t\in\mathbb{N}^{*}:Y_{(1)}(t)-Y_{(2)}(t)\geq c_1\}$. Note that when the event $\mathcal{H}_1$ holds, we have $T_2\leq C_1\log(n+1)$. 

By Proposition \ref{P2.1}, there exists a positive constant $M\in \mathbb{N}$ that depends only on $q,\beta,\delta$, such that for any $t\geq M$ and $\bm{\alpha}\in\mathcal{X}_{c_1}$, $\|\Phi^{(t)}(\bm{\alpha})-\pmb{\alpha}_0\|_1\leq \delta\slash 2$, where $\mathcal{X}_{c_1}$ and $\Phi(\cdot)$ are as in Definition \ref{Def1.1}. We take
\begin{equation*}
    w_1:=\frac{\delta}{10\beta (2\beta q)^M}, \quad w_2:=\frac{(1-\rho_0)\delta}{5 \sqrt{q} \beta^2}, \quad w_0:=\min\{w_1,w_2\},\quad \mathcal{D}_3:=\bigcap_{t=1}^{\lfloor K\log(n+1) \rfloor} \mathcal{E}_{t,w_0,w_0},
\end{equation*}
where $\mathcal{E}_{t,w_0,w_0}$ is as in Definition \ref{def_EA}.
Note that $w_1,w_2,w_0$ depend only on $q,\beta, \delta$. By Lemma \ref{L2.6} and the union bound,
\begin{equation}\label{Eq3.22}
    \mathbb{P}(\mathcal{D}_3^c)\leq 4qK\log(n+1)\exp(-nw_0^2).
\end{equation}

In the following, we assume that the event $\mathcal{H}_1\cap\mathcal{D}_3$ holds. For any $t\in \mathbb{N}$, we let $\gamma_t:=\Phi^{(t)}(s(\mathbf{Y}(T_2)))$. As $s(\mathbf{Y}(T_2))\in\mathcal{X}_{c_1}$, for any $t\geq M$,
\begin{equation}\label{Eq3.20}
    \|\gamma_t-\pmb{\alpha}_0\|_1=\|\Phi^{(t)}(s(\mathbf{Y}(T_2)))-\pmb{\alpha}_0\|_1\leq \frac{\delta}{2}.
\end{equation}
For any $t\in \mathbb{N}$ such that $t\leq K\log(n+1)$, we have
\begin{eqnarray*}
   && \sum_{\ell=1}^q\big(s_{\ell}(\mathbf{Y}(t+1))-\Phi_{\ell}(s(\mathbf{Y}(t)))\big)^2\nonumber\\
    &\leq& 2\sum_{\ell=1}^q\big(s_{\ell}(\mathbf{Y}(t+1))-\Phi_{\ell}(s(\mathbf{Z}(t))) \big)^2+2\sum_{\ell=1}^q\big(\Phi_{\ell}(s(\mathbf{Z}(t)))-\Phi_{\ell}(s(\mathbf{Y}(t)))\big)^2\nonumber\\
   &\leq& 2\sum_{\ell=1}^q\big(\widetilde{Y}_{\ell}(t+1)-\Phi_{\ell}(\mathbf{Z}(t))\big)^2+2\sum_{\ell=1}^q \big(\Phi_{\ell}(\mathbf{Z}(t))-\Phi_{\ell}(\mathbf{Y}(t))\big)^2\nonumber\\
   &\leq& 
   2\sum_{\ell=1}^q\big(\widetilde{Y}_{\ell}(t+1)-\Phi_{\ell}(\mathbf{Z}(t))\big)^2+2\big\|\Phi(\mathbf{Z}(t))-\Phi(\mathbf{Y}(t))\big\|_1^2\nonumber\\
   &\leq& 2\sum_{\ell=1}^q\big(\widetilde{Y}_{\ell}(t+1)-\Phi_{\ell}(\mathbf{Z}(t))\big)^2+8\beta^2 q^2\|\mathbf{Z}(t)-\mathbf{Y}(t)\|_{\infty}^2\leq   2q w_0^2+8\beta^2 q^2 w_0^2\leq 10\beta^4 w_0^2,
\end{eqnarray*}
where $\tilde{Y}_{\ell}(t+1)$ is as in \eqref{Eq3.4} and $s_{\ell}(\mathbf{Y}(t+1))$ is the $\ell$th component of $s(\mathbf{Y}(t+1))$, and we use the rearrangement inequality in the second inequality, Lemma \ref{L1.2} in the fourth inequality, and the fact that we are working on $\mathcal{E}_{t,w_0,w_0}\supseteq \mathcal{D}_3$ in the fifth inequality. Hence
\begin{equation}\label{Eq3.14}
    \|s(\mathbf{Y}(t+1))-\Phi(s(\mathbf{Y}(t)))\|_1\leq \sqrt{q\sum_{\ell=1}^q\big(s_{\ell}(\mathbf{Y}(t+1))-\Phi_{\ell}(s(\mathbf{Y}(t)))\big)^2}\leq 5\sqrt{q}\beta^2  w_0. 
\end{equation}




For any $t\in \mathbb{N}$ such that $T_2+t\leq K\log(n+1)$, by (\ref{Eq3.14}) and Lemma \ref{L1.2},
\begin{eqnarray*}
 && \|s(\mathbf{Y}(T_2+t))-\gamma_t\|_1\nonumber\\
 &\leq& \|s(\mathbf{Y}(T_2+t))-\Phi(s(\mathbf{Y}(T_2+t-1)))\|_1+\|\Phi(s(\mathbf{Y}(T_2+t-1)))-\Phi(\gamma_{t-1})\|_1 \nonumber\\
  &\leq& 5\sqrt{q}\beta^2 w_0+2\beta q \|s(\mathbf{Y}(T_2+t-1))-\gamma_{t-1}\|_1.
\end{eqnarray*}
Hence
\begin{align*}
    \|s(\mathbf{Y}(T_2+t))-\gamma_t\|_1 &\leq \|s(\mathbf{Y}(T_2+t))-\gamma_t\|_1+\frac{5\sqrt{q}\beta^2 w_0}{2\beta q-1} \nonumber\\
    &\leq 2\beta q\bigg(\|s(\mathbf{Y}(T_2+t-1))-\gamma_{t-1}\|_1+\frac{5\sqrt{q}\beta^2 w_0}{2\beta q-1}\bigg).
\end{align*}
By iterating the above inequality, noting that $\gamma_0=s(\mathbf{Y}(T_2))$, we get
\begin{align*}
    \|s(\mathbf{Y}(T_2+t))-\gamma_t\|_1&\leq  (2\beta q)^t \Big(\|s(\mathbf{Y}(T_2))-\gamma_0\|_1+\frac{5\sqrt{q}\beta^2 w_0}{2\beta q-1}\Big)\nonumber\\
    & =\frac{5\sqrt{q}\beta^2}{2\beta q-1}(2\beta q)^tw_0 \leq 5\beta (2\beta q)^t w_0,
\end{align*}
where the last inequality uses $\beta> q\geq 2$. By \eqref{Eq3.20} and the above display, if $T_2+M\leq K\log(n+1)$, then
\begin{equation*}
    \|s(\mathbf{Y}(T_2+M))-\pmb{\alpha}_0\|_1  \leq \|s(\mathbf{Y}(T_2+M))-\gamma_M\|_1+\|\gamma_M-\pmb{\alpha}_0\|_1 \leq 5\beta (2\beta q)^M w_0 +  \frac{\delta}{2} \leq \delta,
\end{equation*}
where the last inequality uses the choice of $M$ and $w_0$.


Now we show by induction that for any $t\in \mathbb{N}$ such that $M\leq t\leq K\log(n+1)-T_2$, we have $\|s(\mathbf{Y}(T_2+t))-\bm{\alpha}_0\|_1\leq \delta$. By the above display, this holds when $t=M$. Now we assume that $M+1\leq t \leq K\log(n+1)-T_2$ and $\|s(\mathbf{Y}(T_2+t-1))-\pmb{\alpha}_0\|_1\leq \delta$. Then, using the fact that $\Phi(\bm{\alpha}_0)=\bm{\alpha}_0$ (see Definition \ref{Def1.2}) and the triangle inequality, we get
\begin{eqnarray*}\label{Eq3.19}
  &&  \|s(\mathbf{Y}(T_2+t))-\pmb{\alpha}_0\|_1=\|s(\mathbf{Y}(T_2+t))-\Phi(\bm{\alpha}_0)\|_1\nonumber\\
  &\leq& \|s(\mathbf{Y}(T_2+t))-\Phi(s(\mathbf{Y}(T_2+t-1)))\|_1+\|\Phi(s(\mathbf{Y}(T_2+t-1)))-\Phi(\bm{\alpha}_0)\|_1\nonumber\\
  &\leq& 5\sqrt{q}\beta^2w_0+\rho_0\|s(\mathbf{Y}(T_2+t-1))-\bm{\alpha}_0\|_1\nonumber\\
  & \leq & (1-\rho_0)\delta+\rho_0\delta \leq \delta,
\end{eqnarray*}
where we use \eqref{Eq3.14} with $t$ replaced by $t+T_2-1$ and Lemma \ref{L1.4} in the third line, and the choice of $w_0$ and the induction hypothesis in the fourth line. This proves the claim for all $t\in \mathbb{N}$ such that $M\leq t\leq K\log(n+1)-T_2$ by induction. 

On the event $\mathcal{H}_1$, we have
\begin{equation}
     T_2+M\leq C_1\log(n+1)+M\leq (C_1+1)\log(n+1),
\end{equation}
where the second inequality holds for all $n$ large enough (depending on $q,\beta,\delta$).
Hence for any $t\in [(C_1+1)\log(n+1),K\log(n+1)]\cap\mathbb{N}$, we have $\mathcal{H}_1\cap \mathcal{D}_3\subseteq \mathcal{V}_{t,\delta}$, and so $(\mathcal{V}_{t,\delta})^c\cap\mathcal{H}_1\subseteq \mathcal{D}_3^c$.
Consequently, by (\ref{Eq3.22}), 
\begin{equation}
    \mathbb{P}((\mathcal{V}_{t,\delta})^c\cap\mathcal{H}_1)\leq \mathbb{P}(\mathcal{D}_3^c)\leq 4qK\log(n+1)\exp(-nw_0^2)\leq C'\exp(-c'n),
\end{equation}
where $C',c'$ are positive constants that depend only on $q,\beta,K,\delta$. 
\end{proof}





\subsection{Proof of Theorem \ref{Thm3.3}}\label{subsec:proof:Thm3.3_part4}

In this subsection, we complete the proof of Theorem \ref{Thm3.3}. Consider any starting state $\mathbf{X}(0)=(X_1(0),\cdots,X_n(0))\in [q]^n$. Let $\mathbf{X}'(0)=(X'_{1}(0),\cdots,X'_{n}(0))$ be drawn from the stationary distribution. For any $t\in \mathbb{N}$, we denote by $\mathbf{X}(t)$ and $\mathbf{X}'(t)$ the states after running $t$ steps of the modified AG sampler started from $\mathbf{X}(0)$ and $\mathbf{X}'(0)$, respectively. For each $t\in\mathbb{N}$ and $\ell\in [q]$, we let 
\begin{equation}\label{YY'}
    Y_{\ell}(t):=n^{-1}\sum_{i=1}^n 1\{X_{i}(t)=\ell\},\qquad Y'_{\ell}(t):=n^{-1}\sum_{i=1}^n 1\{X'_{i}(t)=\ell\}.
\end{equation}

Let $C_1,c_1$ be defined as in Proposition \ref{P2.2}. By a slight abuse of notation (see Definition \ref{def_EA}), for any $t\in\mathbb{N}$, let $\mathcal{F}_t$ be the $\sigma$-algebra generated by $\{\mathbf{X}(s)\}_{s=0}^{t},\{\mathbf{Z}(s)\}_{s=0}^{t-1}$ and $\{\mathbf{X}'(s)\}_{s=0}^{t},\{\mathbf{Z}'(s)\}_{s=0}^{t-1}$, and let $\mathcal{G}_t$ be the $\sigma$-algebra generated by $\{\mathbf{X}(s)\}_{s=0}^{t},\{\mathbf{Z}(s)\}_{s=0}^{t}$ and $\{\mathbf{X}'(s)\}_{s=0}^{t}$, $\{\mathbf{Z}'(s)\}_{s=0}^{t}$. Let $\rho_0,\delta_0$ be the constants in Lemma \ref{L1.4}; note that $\rho_0,\delta_0$ only depend on $q,\beta$. 

Below we construct a coupling between $\mathbf{X}(t)$ and $\mathbf{X}'(t)$ for all $t\in\mathbb{N}$. For any $t\in \mathbb{N}$, given $\mathbf{X}(t)$ and $\mathbf{X}'(t)$, we couple $\mathbf{X}(t+1)$ and $\mathbf{X}'(t+1)$ as follows. Assume that $Y_{\ell_1}(t)\geq Y_{\ell_2}(t)\geq \cdots \geq Y_{\ell_q}(t)$ and $Y_{\ell_1'}'(t)\geq Y_{\ell_2'}'(t)\geq \cdots\geq Y_{\ell_q'}'(t)$ (where $(\ell_1,\cdots,\ell_q)$ and $(\ell_1', \cdots, \ell_q')$ are permutations of $[q]$). Let $\lambda\in S_q$ be such that $\lambda(\ell_k)=\ell_{k}'$ for each $k\in [q]$. For each $\ell\in [q]$, we independently sample $W_{\ell}(t)\sim N(0,(n\beta)^{-1})$. Let
\begin{equation}\label{WZZ}
    W'_{\ell}(t)=W_{\lambda^{-1}(\ell)}(t), \qquad Z_{\ell}(t)=Y_{\ell}(t)+W_{\ell}(t), \qquad Z'_{\ell}(t)=Y'_{\ell}(t)+W'_{\ell}(t).
\end{equation}
For each $i\in [n]$, we sample $(\widetilde{X}_{i}(t+1),\widetilde{X}'_{i}(t+1))\in [q]^2$ such that for each $\ell\in [q]$,
\begin{equation*}
        \mathbb{P}(\widetilde{X}_{i}(t+1)=\ell|\mathcal{G}_t)=\frac{e^{\beta  Z_{\ell}(t)}}{\sum_{\ell'=1}^q e^{\beta Z_{\ell'}(t)}}, \quad  \mathbb{P}(\widetilde{X}'_{i}(t+1)=\ell|\mathcal{G}_t)=\frac{e^{\beta  Z'_{\ell}(t)}}{\sum_{\ell'=1}^q e^{\beta Z'_{\ell'}(t)}},
\end{equation*}
and $\mathbb{P}(\widetilde{X}_{i}(t+1)\neq \lambda^{-1}(\widetilde{X}'_{i}(t+1))|\mathcal{G}_t)$ is equal to the total variation distance between the conditional laws of $\widetilde{X}_{i}(t+1)$ and $\lambda^{-1}(\widetilde{X}'_{i}(t+1))$ given $\mathcal{G}_t$. We sample $\tau_{t}$ uniformly from $S_q$, and let $\tau_{t}'=\tau_{t}\lambda^{-1}\in S_q$. Finally, set $X_{i}(t+1)=\tau_{t}(\widetilde{X}_{i}(t+1))$ and $X'_{i}(t+1)=\tau'_{t}(\widetilde{X}'_{i}(t+1))$. 

We fix $K\geq 2(C_1+1)$, whose value will be determined later. We also let $\delta=\delta_0\slash 2$ and
\begin{align*}
    t_1:=\lfloor C_1\log(n+1)\rfloor,\qquad t_2:=\lceil (C_1+1)\log(n+1)\rceil, \qquad   t_3:=\lfloor K\log(n+1)\rfloor.
\end{align*}
By another slight abuse of notation (see Proposition \ref{P2.2}), let $\mathcal{H}_1$ be the event that $Y_{(1)}(t)-Y_{(2)}(t)\geq c_1$ for some $t\in [t_1]$ and $Y'_{(1)}(t')-Y'_{(2)}(t')\geq c_1$ for some $t'\in [t_1]$. Note that $\mathcal{H}_1\in\mathcal{F}_{t_1}$, and by Proposition \ref{P2.2}, 
\begin{equation}\label{boundonH}
    \mathbb{P}(\mathcal{H}_1^c)\leq C\exp(-c\log\log(n+1)). 
\end{equation}

For any $t\in \mathbb{N}$, let $\mathcal{V}_{t,\delta}$ be the event that $\|s(\mathbf{Y}(t))-\bm{\alpha}_0\|_1\leq \delta$ and $\|s(\mathbf{Y}'(t))-\bm{\alpha}_0\|_1\leq \delta$ (where $s(\mathbf{Y}(t))$ and $s(\mathbf{Y}'(t))$ are the order statistics of $\mathbf{Y}(t)$ and $\mathbf{Y}'(t)$, respectively), and let $\mathcal{U}_{t,\delta}$ be the event that $\max_{\ell\in [q]}\{|W_{\ell}(t)|\}\leq \delta\slash q$. Let $\mathcal{W}_{K,\delta}:=\bigcap_{t\in [t_2,t_3]\cap\mathbb{N}}\mathcal{V}_{t,\delta}$. 
By Proposition \ref{P2.3} and tail bounds for Gaussian random variables, we have  
\begin{equation}\label{Eq3.33}
    \mathbb{P}((\mathcal{W}_{K,\delta})^c\cap \mathcal{H}_1)\leq C'\exp(-c' n),\quad\mathbb{P}((\mathcal{U}_{t,\delta})^c)\leq 2q\exp(-n\beta  q^{-2}\delta^2 \slash 2) \leq C'\exp(-c' n),
\end{equation}
where $C',c'$ are positive constants that only depend on $q,\beta,K$. 
Let $H(\cdot,\cdot)$ be the Hamming distance on $[q]^n$. For any $t\in \mathbb{N}$,
\begin{equation*}
    H(\mathbf{X}(t+1),\mathbf{X}'(t+1))=\sum_{i=1}^n 1\{X_{i}(t+1)\neq X'_{i}(t+1)\}=\sum_{i=1}^n 1\{\widetilde{X}_{i}(t+1)\neq\lambda^{-1}(\widetilde{X}'_{i}(t+1))\},
\end{equation*}
hence
\begin{eqnarray}\label{Eq3.31}
     \mathbb{E}[H(\mathbf{X}(t+1),\mathbf{X}'(t+1))|\mathcal{F}_t]&=&\sum_{i=1}^n \mathbb{P}(\widetilde{X}_{i}(t+1)\neq \lambda^{-1}(\widetilde{X}'_{i}(t+1))|\mathcal{F}_t)\nonumber\\
    &=& \frac{1}{2}\sum_{i=1}^n\sum_{\ell=1}^q \mathbb{E}\bigg[\bigg|\frac{e^{\beta  Z_{\ell}(t)}}{\sum_{\ell'=1}^q e^{\beta Z_{\ell'}(t)}} - \frac{e^{\beta  Z'_{\lambda(\ell)}(t)}}{\sum_{\ell'=1}^q e^{\beta Z'_{\ell'}(t)}}\bigg|\bigg|\mathcal{F}_t\bigg],
\end{eqnarray}
where the last equality uses the fact that $(\widetilde{X}_{i}(t+1),\lambda^{-1}(\widetilde{X}'_{i}(t+1)))$ are coupled using the total variation coupling. Note that when $\mathcal{U}_{t,\delta}\cap\mathcal{V}_{t,\delta}$ holds, we have
\begin{eqnarray*}
 && \|(Z_{\ell_1}(t),\cdots,Z_{\ell_q}(t))-\bm{\alpha}_0\|_1\nonumber\\
 &\leq& \|(Z_{\ell_1}(t),\cdots,Z_{\ell_q}(t))-(Y_{\ell_1}(t),\cdots,Y_{\ell_q}(t))\|_1+\|s(\mathbf{Y}(t))-\bm{\alpha}_0\|_1\nonumber\\
  &\leq& q\max_{\ell\in [q]}\{|W_{\ell}(t)|\}+\delta\leq 2\delta\leq \delta_0.
\end{eqnarray*}
The same bound applies to $(Z'_{\ell_1'}(t),\cdots,Z'_{\ell_q'}(t))$, and consequently, on the set $\mathcal{U}_{t,\delta}\cap\mathcal{V}_{t,\delta}$,
\begin{eqnarray}\label{Eq3.32}
     \sum_{\ell=1}^q\bigg|\frac{e^{\beta  Z_{\ell}(t)}}{\sum_{\ell'=1}^q e^{\beta Z_{\ell'}(t)}} - \frac{e^{\beta  Z'_{\lambda (\ell)}(t)}}{\sum_{\ell'=1}^q e^{\beta Z'_{\ell'}(t)}}\bigg|
    &\leq& \rho_0\sum_{\ell = 1}^q|Z_{\ell}(t)-Z'_{\lambda(\ell)}(t)|=\rho_0\sum_{\ell=1}^q|Y_{\ell}(t)-Y'_{\lambda(\ell)}(t)|\nonumber\\
    &\leq& \rho_0\sum_{\ell=1}^q|Y_{\ell}(t)-Y'_{\ell}(t)|\leq\frac{2}{n}\rho_0 H(\mathbf{X}(t),\mathbf{X}'(t)),
\end{eqnarray}
where the inequality in the first line uses
Lemma \ref{L1.4}, the equality in the first line uses \eqref{WZZ}, the first inequality in the second line uses the fact that $(Y_{\ell}(t))_{\ell\in[q]}$ and $(Y'_{\lambda(\ell)}(t))_{\ell\in[q]}$ have the same order, and the second inequality in the second line uses \eqref{YY'}. 
By \eqref{Eq3.31} and \eqref{Eq3.32}, we have
\begin{equation*}
    \mathbb{E}[H(\mathbf{X}(t+1),\mathbf{X}'(t+1))|\mathcal{F}_t]\leq \rho_0 H(\mathbf{X}(t),\mathbf{X}'(t))+\frac{nq}{2}(\mathbb{P}((\mathcal{U}_{t,\delta})^c|\mathcal{F}_t)+\mathbb{P}((\mathcal{V}_{t,\delta})^c|\mathcal{F}_t)).
\end{equation*}
For any $t\in [t_2,t_3]\cap\mathbb{N}$, by \eqref{Eq3.33}, noting that $\mathcal{H}_1\in\mathcal{F}_{t_1}\subseteq \mathcal{F}_t$, we have 
\begin{eqnarray*}
   && \mathbb{E}[H(\mathbf{X}(t+1),\mathbf{X}'(t+1))\mathbbm{1}_{\mathcal{H}_1}]=\mathbb{E}[\mathbb{E}[H(\mathbf{X}(t+1),\mathbf{X}'(t+1))|\mathcal{F}_t]\mathbbm{1}_{\mathcal{H}_1}]\nonumber\\
    &\leq& \rho_0\mathbb{E}[H(\mathbf{X}(t),\mathbf{X}'(t))\mathbbm{1}_{\mathcal{H}_1}]+\frac{nq}{2}(\mathbb{P}((\mathcal{U}_{t,\delta})^c)+\mathbb{P}((\mathcal{V}_{t,\delta})^c\cap\mathcal{H}_1))\nonumber\\
    &\leq& \rho_0\mathbb{E}[H(\mathbf{X}(t),\mathbf{X}'(t))\mathbbm{1}_{\mathcal{H}_1}]+\frac{nq}{2}(\mathbb{P}((\mathcal{U}_{t,\delta})^c)+\mathbb{P}((\mathcal{W}_{K,\delta})^c\cap\mathcal{H}_1))\nonumber\\
    &\leq& \rho_0\mathbb{E}[H(\mathbf{X}(t),\mathbf{X}'(t))\mathbbm{1}_{\mathcal{H}_1}]+C'\exp(-c' n).
\end{eqnarray*}
Hence
\begin{equation*}
    \mathbb{E}[H(\mathbf{X}(t+1),\mathbf{X}'(t+1))\mathbbm{1}_{\mathcal{H}_1}]-\frac{C'\exp(-c'n)}{1-\rho_0}\leq \rho_0 \bigg(  \mathbb{E}[H(\mathbf{X}(t),\mathbf{X}'(t))\mathbbm{1}_{\mathcal{H}_1}]-\frac{C'\exp(-c'n)}{1-\rho_0}\bigg),
\end{equation*}
and consequently,
\begin{align*}
    \mathbb{E}[H(\mathbf{X}(t_3),\mathbf{X}'(t_3))\mathbbm{1}_{\mathcal{H}_1}]-\frac{C'\exp(-c'n)}{1-\rho_0}&\leq\rho_0^{t_3-t_2}\bigg(\mathbb{E}[H(\mathbf{X}(t_2),\mathbf{X}'(t_2))\mathbbm{1}_{\mathcal{H}_1}]-\frac{C'\exp(-c'n)}{1-\rho_0}\bigg)\nonumber\\
    &\leq \rho_0^{t_3-t_2} n\leq \rho_0^{K\log(n+1)\slash 3}n,
\end{align*}
where the last inequality holds for all $n$ large enough (depending on $q,\beta, K$). Taking $K$ sufficiently large (depending on $q,\beta$) and $n$ sufficiently large (depending on $q,\beta,K$), by \eqref{boundonH} and the above display, we get
\begin{equation}
    \mathbb{P}(\mathcal{H}_1^c)\leq \frac{1}{8}, \qquad \mathbb{P}(\{\mathbf{X}(t_3)\neq \mathbf{X}'(t_3)\}\cap\mathcal{H}_1)\leq \mathbb{E}[H(\mathbf{X}(t_3),\mathbf{X}'(t_3))\mathbbm{1}_{\mathcal{H}_1}]\leq \frac{1}{8}.
\end{equation}
Hence $\mathbb{P}(\mathbf{X}(t_3)\neq \mathbf{X}'(t_3))\leq \frac{1}{4}$.
Thus by the coupling lemma \cite[Theorem 5.4]{MR3726904}, $t_{\mathrm{mix}}\Big(\frac{1}{4}\Big)\leq t_3\leq C\log(n+1)$.
By \cite[(4.34)]{MR3726904}, we obtain the conclusion of the theorem.

\section{Discussion}

This paper builds on a growing literature on sampling methods for Ising and Potts models, using an auxiliary Gaussian variable, a strategy first proposed in the machine learning literature by \citet{Martens:2010}.
We present two novel choices of auxiliary Gaussian variables: (i) a choice for Potts models, which leads to a sampler whose cost per iteration scales linearly in $q$; and (ii) a choice which automatically detects the (approximately) low-rank structure of $\bf A$ and takes advantage of this low-rank structure, if it exists.
We primarily focus on the AG sampler, which is a block Gibbs sampler.
We run an extensive suite of numerical experiments and identify many regimes where the AG sampler outperforms competitors, at times yielding an ESS/s which is orders of magnitudes higher---for example for Ising models on a graph and low-rank Hopfield networks. 
We also identify regimes where the AG sampler exhibits poor performance, for example when applied to a cold spin glass system.
Overall, we believe that the AG sampler's ease of implementation and competitive performance make it an excellent method to sample from Potts models.

Our paper also provides bounds on the mixing time of AG samplers.
Such bounds exist for the Heat Bath and Swendsen-Wang algorithms.
To the best of our knowledge, our paper provides the first bounds for AG samplers. Our bounds are of similar order as those for Swendsen-Wang algorithms (see, e.g., \cite{Galanis:2019}).

There exist several directions for future research. First, it would be of interest to establish matching lower bounds on the mixing time for Theorems~\ref{Thm3.1} and~\ref{Thm3.3}. Second, it would be interesting to extend Theorem~\ref{Thm3.3} to the low-temperature regime for broader classes of models, such as Potts models on random regular graphs. Another direction for future research is to investigate other algorithms which use the auxiliary Gaussian variables we propose.
This includes discrete HMC \citep{Zhang:2012} and more generally gradient-based methods to sample over $\bf Z$.
An important challenge here is to understand the geometry of $p({\bf Z})$ and which methods---or which tuning of a method---enable efficient sampling.
Similarly, it is of interest to study how tempering best works with an AG sampler, or any sampler over $\mathbb P({\bf X}, {\bf Z})$ in order to sample from cold systems.

\section*{Acknowledgments}

SM gratefully acknowledges NSF for partial support during this research (DMS-2113414, DMS-2515519).

\appendix

\section*{Appendix}

The Appendix contains proofs missing in the main body.
We derive the conditional and marginal distributions, that arise when we introduce either a regular or low-rank auxiliary Gaussian variable for Potts models.
We prove Proposition~\ref{thm:low-rank}, which characterizes the error committed in using the low-rank algorithm, and Proposition~\ref{prop:low_rank}, which characterizes the type of coupling matrices for which we may want to use a low-rank algorithm.
Specialized algorithms for the Ising model ($q = 2$) are derived. We also provide an overview on auxiliary cluster and tempering algorithms.

The code used for our numerical experiments can be found at \url{https://github.com/charlesm93/potts_simulation}.
The ReadMe provides instructions on how to run the code.

\section{Missing proofs in Section \ref{sec:choice}}

\subsection{Conditional and marginal distributions when using the regular Auxiliary Gaussian}\label{sec:choice:1}

We begin by proving Equations~\eqref{eq:conditional} and \eqref{eq:marginal}.
Recall from \eqref{eq:potts_simplified} that the marginal p.m.f.~of $\mathbf X$ under $\mathbb P$ is proportional to
\[\exp\Big(\frac{1}{2}\sum_{\ell=1}^q {\bf y}_\ell'B{\bf y}_\ell\Big).\]
Also from \eqref{eq:main}, the conditional density of $\mathbf Z$ given $\mathbf X=\mathbf x$ is proportional to
\[\exp\Big(-\frac{1}{2}\sum_{\ell=1}^q (\mathbf z_\ell-\mathbf y_\ell)'B (\mathbf z_\ell-\mathbf y_\ell)\Big).\]
Thus, the joint distribution of $\mathbf X$ and $\mathbf Z$ is proportional to
\begin{align}\label{eq:joint1}
&\notag\exp\left(\frac{1}{2}\sum_{\ell=1}^q \Big[{\bf y}_\ell'B{\bf y}_\ell-(\mathbf z_\ell-\mathbf y_\ell)'B (\mathbf z_\ell-\mathbf y_\ell)\Big]\right)\\
\notag=&\exp\left(-\frac{1}{2}\sum_{\ell=1}^q \mathbf z_\ell'B\mathbf z_\ell+\sum_{\ell=1}^q \mathbf z_\ell'B\mathbf y_\ell\right)\\
=&\exp\left(-\frac{1}{2}\sum_{\ell=1}^q \mathbf z_\ell'B\mathbf z_\ell+\sum_{\ell=1}^q\sum_{i=1}^n \sum_{j=1}^n z_{\ell j} B_{ij} 1\{x_i=\ell\}\right).
\end{align}
From \eqref{eq:joint1}, we see that given $\mathbf Z$, the random variables $(X_1,\ldots,X_n)$ are mutually independent, with
$$\mathbb{P}(X_i=\ell|\mathbf Z) \propto \exp\Big(\sum_{j=1}^nB_{ij} z_{\ell j}\Big),$$
which verifies \eqref{eq:conditional}.

The marginal distribution of $\mathbf Z$ is obtained from \eqref{eq:joint1} by summing over $\mathbf x\in [q]^n$, from which \eqref{eq:marginal} follows.

\subsection{Conditional and marginal distributions when using the low-rank Auxiliary Gaussian}\label{sec:choice:2}

Next we prove Equations~\eqref{eq:main2} and \eqref{eq:low_rank_marginal}.
Recall from \eqref{eq:low_rank_main} that the marginal p.m.f.~of $\mathbf X$ under $\mathbb P$ is proportional to
$$\exp\Big(\frac{1}{2}\sum_{\ell=1}^q \sum_{j=1}^k \mu_j (\mathbf p_j'y_\ell)^2\Big).$$
Also from \eqref{eq:main2}, the conditional density of $\mathbf Z$ given $\mathbf X=\mathbf x$ is proportional to
$$\exp\Big(-\frac{1}{2}\sum_{\ell=1}^q \sum_{j=1}^k \mu_j(z_{\ell j}-\mathbf p_j'\mathbf y_\ell)^2\Big).$$
Thus, the joint distribution of $\mathbf X$ and $\mathbf Z$ is proportional to
\begin{align}\label{eq:joint2}
&\notag\exp\left(\frac{1}{2}\sum_{\ell=1}^q \sum_{j=1}^k\mu_j\Big[ (\mathbf p_j'y_\ell)^2-(z_{\ell j}-\mathbf p_j'\mathbf y_\ell)^2\Big]\right)\\
\notag=&\exp\Big(-\frac{1}{2}\sum_{\ell=1}^q \sum_{j=1}^k \mu_j z_{\ell j}^2+\sum_{\ell=1}^q\sum_{j=1}^k \mu_j z_{\ell j} \mathbf p_j'\mathbf y_\ell \Big)\\
=&\exp\Big(-\frac{1}{2}\sum_{\ell=1}^q \sum_{j=1}^k \mu_j z_{\ell j}^2+\sum_{\ell=1}^q\sum_{j=1}^k\sum_{i=1}^n \mu_j z_{\ell j} p_{j i} 1\{x_i=\ell\} \Big).
\end{align}
Using \eqref{eq:joint2},  given $\mathbf Z=\mathbf z$ the random variables $(X_1,\ldots,X_n)$ are mutually independent, with
$$\mathbb{P}(X_i=\ell|\mathbf Z) \propto \exp\Big(\sum_{j=1}^k \mu_j z_{\ell j} p_{ji}\Big),$$
which verifies \eqref{eq:low_rank_conditional}. The marginal distribution of $\mathbf Z$ is obtained from \eqref{eq:joint2} by summing over $\mathbf x\in [q]^n$, from which \eqref{eq:low_rank_marginal} follows.

\subsection{Proof of Proposition~\ref{thm:low-rank}}\label{sec:choice:3}

With $B$ and $\tilde{B}$ as in the distributions $\mathbb P$  and $\mathbb Q$ in \eqref{eq:potts_simplified}
and \eqref{eq:potts_low_rank} respectively, for any $\mathbf x\in [q]^n$, we have
\begin{align}
\notag\Big|\sum_{i,j=1}^nB_{ij}1\{x_i=x_j\}-\sum_{i,j=1}^n \tilde{B}_{ij}1\{x_i=x_j\}\Big|=&\Big| \sum_{\ell=1}^q \mathbf y_\ell'(B-\tilde B)\mathbf y_\ell\Big|\\
\notag\le &\sum_{\ell=1}^q \| \mathbf y_\ell\|_2^2 \|B-\tilde B\|_2\\
=& \|B-\tilde B\|_2\sum_{\ell=1}^q \sum_{i=1}^n 1\{x_i=\ell\}\\
=&n \|B-\tilde B\|_2\le  n\varepsilon  ,
\end{align}
where the last bound uses the definition of $\tilde B$ to note that all eigenvalues of $B-\tilde B$ are smaller than $\varepsilon$ in absolute value.
Thus, for any $\mathbf x\in [q]^n$, we have
\begin{align}\label{eq:pmf_ratio}
e^{\frac{-n\beta\varepsilon}{2}}\le \frac{\exp\Big(\frac{\beta}{2}\sum_{i,j=1}^nB_{ij}1\{x_i=x_j\}\Big)}{\exp\Big(\frac{\beta}{2}\sum_{i,j=1}^n\tilde B_{ij}1\{x_i=x_j\}\Big)}\le e^{\frac{n\beta\varepsilon}{2}}.
\end{align}

On summing over $\mathbf x\in [q]^n$, this gives
\begin{align}\label{eq:part_ratio}
e^{-\frac{n\beta\varepsilon}{2}}\le \frac{Z(\beta,B)}{Z(\beta,\tilde{B})}\le e^{\frac{n\beta\varepsilon}{2}},
\end{align}
which verifies part (i). For verifying part (ii), taking a ratio of \eqref{eq:pmf_ratio} and \eqref{eq:part_ratio} we get
\begin{align*}
e^{-n\beta\varepsilon}\le \frac{ \mathbb P(\mathbf X=\mathbf x)}{\mathbb Q(\mathbf X=\mathbf x)}\le e^{n\beta\varepsilon},
\end{align*}
which gives
$\max({\rm KL}(\mathbb P|\mathbb Q),{\rm KL}(\mathbb Q|\mathbb P))\le n\beta \varepsilon$. This verifies part (ii), and hence completes the proof of the lemma.

\subsection{Proof of Proposition \ref{prop:low_rank}}\label{sec:choice:4}

To begin, we bound  $k$ by examining the trace of $B^2$, and noting that, per Markov's inequality,
\begin{equation*}
  k \le \sum_{i = 1}^n \mu_i^2 / \varepsilon^2 = {\rm tr}(B^2) / \varepsilon^2.
\end{equation*}
Also, we have
\begin{align}\label{eq:bound}
{\rm tr}(B^2)=\beta^2{\rm tr}(A^2+2|\lambda_{\min}| A+\lambda_{\min}^2 I)=\beta^2{\rm tr}(A^2)+\beta^2 n\lambda_{\min}^2.
\end{align}
Combining the above two displays, it suffices to show that the RHS of \eqref{eq:bound} is $o(n)$.
Since $\lambda_{\min}=o(1)$, it suffices to show ${\rm tr}(A^2)=o(n)$, which is the focus of the rest of the proof.

To this effect, recall that ${\rm tr}(A)=0$. Thus, with $\lambda_1(A),\ldots,\lambda_n(A)$ denoting the eigenvalues of $A$ we have
\[\sum_{i=1}^n\lambda_i(A) 1\{\lambda_i(A)>0\}=\sum_{i=1}^n |\lambda_i(A)|1\{\lambda_i(A)<0\}\le n |\lambda_{\min}|.\]
This in turn gives 
\[\sum_{i=1}^n|\lambda_i(A)|=\sum_{i=1}^n\lambda_i(A) 1\{\lambda_i(A)>0\}+\sum_{i=1}^n|\lambda_i(A)| 1\{\lambda_i(A)<0\}\le 2n|\lambda_{\min}|=o(n),\]
and so
\[{\rm tr}(A^2)=\sum_{i=1}^n\lambda_i(A)^2\le \max\{|\lambda_{\max}|,|\lambda_{\min}|\} \sum_{i=1}^n|\lambda_i(A)|=o(n),\]
where the last equality uses the assumption that $\lambda_{\max} =O(1)$ and $\lambda_{\min}=o(1)$. This completes the proof of the proposition.

\section{Missing proofs in Section \ref{sec:proof2}}\label{secB}
\subsection{Proof of Lemma \ref{L2.1}}\label{Proofs:L2.1}

For any $x\in [0,1]$, let $\psi(x):=(1-x)e^{\beta x}-(q-1)x-1$, and note that
\begin{align*}
    \psi(0)=0,\qquad \psi'(x)=-e^{\beta x}+\beta e^{\beta x}(1-x)-(q-1),\qquad \psi''(x)=\beta e^{\beta x} (\beta(1-x)-2).
\end{align*}
Then $\psi''(x)$ has exactly one root in $[0,1]$, and so $\psi'(x)$ has at most two roots in $[0,1]$. As $\psi'(0)=\beta-q>0$ and $\psi'(1)=-e^{\beta}-q+1<0$, there exists a unique root for $\psi'(x)$ in $(0,1)$, which we denote by $x_1$. Since $\psi(x)$ is strictly increasing on $[0,x_1)$ with $\psi(x_1)>\psi(0)=0$, and strictly decreasing on $(x_1,1]$ with $\psi(1)=-q<0$, it follows that there exists a unique $x_0\in (x_1,1)$, such that $\psi(x_0)=0$. Also, recalling the formulas of $\psi,g$, we can write $\psi(x)=(e^{\beta x}-1)\big(1-\frac{x}{g(x)}\big)$. Consequently, this gives 
\begin{align}\label{g_prop}
    g(0)=0;\quad g(x_0)=x_0;\quad g(x)>x\text{ for } x\in (0,x_0);\quad g(x)<x\text{ for }x\in (x_0,1].
\end{align}



We now give the proof of part (a). Suppose $(x,y)$ is an arbitrary solution of \eqref{Eq2.3}, if it exists. Then we have $x-y\in (0,1)$ and  
\begin{equation*}
    g(x-y)=\frac{e^{\beta(x-y)}-1}{e^{\beta(x-y)}+q-1}=\frac{x\slash y-1}{x\slash y+q-1}=\frac{x-y}{x+(q-1)y}=x-y.
\end{equation*}
Since $g$ has a unique fixed point $x_0$ in $(0,1)$, we have $x-y=x_0$. As $x+(q-1)y=1$, solving this we get $x=(1+(q-1)x_0)\slash q$ and $y=(1-x_0)\slash q$. Thus there is at most a unique solution to \eqref{Eq2.3}. The fact that $x=(1+(q-1)x_0)\slash q$ and $y=(1-x_0)\slash q$ is a solution can be verified by elementary computations. As $a-b=x_0>x_1$, by the above analysis, $\psi'(a-b)<0$, which yields $-e^{\beta(a-b)}+\beta e^{\beta(a-b)}(1-a+b)-(q-1)<0$. This, along with \eqref{Eq2.3}, gives
\begin{align*}
   0> -\frac{a}{b}+\beta\cdot \frac{a}{b}\cdot qb-(q-1)=-\frac{a}{b}+\beta aq-(q-1)\Rightarrow \beta q ab<a+(q-1)b=1.
\end{align*}
This completes the proof of part (a).


Part (b) now follows from part (a) and \eqref{g_prop}, on noting that $a-b=x_0$.



Turning to the proof of part (c),
note that for any $x\in [q^{-1},1]$,
\begin{align*}
   & h(x)>x\Leftrightarrow (1-x)e^{\beta x}>x(q-1)e^{\beta\frac{1-x}{q-1}}\Leftrightarrow (1-x)e^{\beta\frac{qx-1}{q-1}}>x(q-1),
\end{align*}
{\small
\begin{align*}
    g\bigg(\frac{qx-1}{q-1}\bigg)>\frac{qx-1}{q-1} \Leftrightarrow (q-1)\Big(e^{\beta\frac{qx-1}{q-1}}-1\Big)>(qx-1)\Big(e^{\beta\frac{qx-1}{q-1}}+q-1\Big)\Leftrightarrow (1-x)e^{\beta\frac{qx-1}{q-1}}>x(q-1).
\end{align*}
}Thus we conclude that $h(x)>x$ if and only if $g\Big(\frac{qx-1}{q-1}\Big)>\frac{qx-1}{q-1}$. When $x\in (q^{-1},a)$, we have  $\frac{qx-1}{q-1}\in \big(0,\frac{qa-1}{q-1}\big)=(0,a-b)$; using part (b) and the above conclusion, we have $h(x)>x$ when $x\in (q^{-1},a)$. Similarly, we can deduce that $h(a)=a$ and $h(x)<x$ when $x\in (a,1]$. The fact that $h(q^{-1})=q^{-1}$ follows from direct computation.


\subsection{Proof of Lemma \ref{L1.2}}

For any $\ell\in [q]$ and $\theta\in [0,1]$, let
\begin{equation}\label{def_p}
    \phi_{\ell}(\theta)=\Phi_{\ell}(\theta\bm{\alpha}+(1-\theta)\bm{\alpha}')=\frac{e^{\beta(\theta\alpha_{\ell}+(1-\theta)\alpha_{\ell}')}}{\sum_{\ell'=1}^q e^{\beta(\theta\alpha_{\ell'}+(1-\theta)\alpha_{\ell'}')}}.
\end{equation}
Note that
\begin{equation*}
    \phi_{\ell}'(\theta)=\frac{\beta e^{\beta(\theta\alpha_{\ell}+(1-\theta)\alpha'_{\ell})}\sum_{\ell'\in [q]\backslash\{\ell\}}e^{\beta(\theta\alpha_{\ell'}+(1-\theta)\alpha'_{\ell'})}\big((\alpha_{\ell}-\alpha_{\ell}')-(\alpha_{\ell'}-\alpha'_{\ell'})\big)}{\big(\sum_{\ell'=1}^q e^{\beta(\theta\alpha_{\ell'}+(1-\theta)\alpha_{\ell'}')}\big)^2}.
\end{equation*}
Hence $|\phi'_{\ell}(\theta)|\leq 2\beta\max_{\ell'\in[q]} |\alpha'_{\ell'}-\alpha_{\ell'}|$.
Therefore, 
\begin{equation*}
    \|\Phi(\bm{\alpha'})-\Phi(\bm{\alpha})\|_1=  \sum_{\ell\in [q]}|\phi_{\ell}(1)-\phi_{\ell}(0)|\leq 2\beta q \max_{\ell'\in [q]} |\alpha'_{\ell'}-\alpha_{\ell'}|.
\end{equation*}

\subsection{Proof of Lemma \ref{L1.4}}

For any $\bm{\alpha}=(\alpha_1,\cdots,\alpha_q)\in\mathbb{R}^q$,
\begin{equation*}
    \frac{\partial \Phi_{\ell}(\bm{\alpha})}{\partial \alpha_{\ell}} = \frac{\beta e^{\beta \alpha_{\ell}}\big(\sum_{s\in [q]\backslash\{\ell\}}e^{\beta \alpha_{s}}\big)}{\big(\sum_{s=1}^q e^{\beta \alpha_{s}}\big)^2}, \quad\forall\ell\in [q],
\end{equation*}
\begin{equation*}
    \frac{\partial \Phi_{\ell}(\bm{\alpha})}{\partial \alpha_{\ell'}}=\frac{-\beta e^{\beta \alpha_{\ell}} e^{\beta \alpha_{\ell'}}}{\big(\sum_{s=1}^q e^{\beta \alpha_{s}}\big)^2},\quad \forall \ell,\ell'\in [q] \text{ such that } \ell\neq \ell'. 
\end{equation*}
In particular, with $\pmb{\alpha}_0$ as in Definition~\ref{Def1.2}, we have
\begin{equation}\label{Eqq3}
    \frac{\partial \Phi_1(\bm{\alpha}_0)}{\partial \alpha_1}=\frac{\beta (q-1) e^{\beta a} e^{\beta b}}{(e^{\beta a}+(q-1) e^{\beta b})^2}, 
\end{equation}
\begin{equation}\label{Eqq2}
    \frac{\partial \Phi_{\ell}(\bm{\alpha}_0)}{\partial \alpha_{\ell}}=\frac{\beta e^{\beta b} (e^{\beta a}+(q-2)e^{\beta b})}{(e^{\beta a}+(q-1) e^{\beta b})^2},\quad   \forall \ell\in [q]\backslash \{1\},
\end{equation}
\begin{equation}
    \frac{\partial \Phi_{\ell}(\bm{\alpha}_0)}{\partial \alpha_{\ell'}}=\frac{-\beta e^{\beta  a} e^{\beta   b}}{(e^{\beta a}+(q-1) e^{\beta b})^2},\quad \forall \ell,\ell'\in [q] \text{ such that } \ell\neq \ell'\text{ and }\min\{\ell,\ell'\}=1, 
\end{equation}
\begin{equation}\label{Eqq1}
    \frac{\partial \Phi_{\ell}(\bm{\alpha}_0)}{\partial \alpha_{\ell'}}=\frac{-\beta e^{2\beta b}}{(e^{\beta a}+(q-1) e^{\beta b})^2},\quad \forall \ell,\ell'\in [q] \text{ such that } \ell\neq \ell'\text{ and }\min\{\ell,\ell'\}\geq 2. 
\end{equation}

For any $\ell,i,j\in [q]$ such that $i<j$, we bound $\sum_{\ell=1}^q \big|\frac{\partial \Phi_{\ell}(\bm{\alpha}_0)}{\partial \alpha_i}-\frac{\partial \Phi_{\ell}(\bm{\alpha}_0)}{\partial \alpha_j}\big|$ as follows. If $i=1$, then $j\in [q]\backslash \{1\}$, and using \eqref{Eqq3}--\eqref{Eqq1}, we get
\begin{align}\label{Eq2.15}
    &\sum_{\ell=1}^q \bigg|\frac{\partial \Phi_{\ell}(\bm{\alpha}_0)}{\partial \alpha_i}-\frac{\partial \Phi_{\ell}(\bm{\alpha}_0)}{\partial \alpha_j}\bigg|\nonumber\\
    =\,&\bigg|\frac{\partial \Phi_{1}(\bm{\alpha}_0)}{\partial \alpha_1}-\frac{\partial \Phi_1(\bm{\alpha}_0)}{\partial \alpha_j}\bigg|+\bigg|\frac{\partial\Phi_j(\bm{\alpha}_0)}{\partial \alpha_1}-\frac{\partial \Phi_j(\bm{\alpha}_0)}{\partial \alpha_j}\bigg|+\sum_{\ell\in[q]\backslash \{1,j\}}\bigg|\frac{\partial \Phi_{\ell}(\bm{\alpha}_0)}{\partial \alpha_1}-\frac{\partial \Phi_{\ell}(\bm{\alpha}_0)}{\partial \alpha_j}\bigg|\nonumber\\
    =\,&\frac{|\beta (q-1) e^{\beta a} e^{\beta b}+\beta e^{\beta a}e^{\beta b}|+|-\beta e^{\beta a}e^{\beta b}-\beta e^{\beta b}(e^{\beta a}+(q-2)e^{\beta b})|}{(e^{\beta a}+(q-1) e^{\beta b})^2}\nonumber\\
    &+\frac{(q-2)|-\beta e^{\beta a}e^{\beta b}+   \beta e^{ 2 \beta b}|}{(e^{\beta a}+(q-1) e^{\beta b})^2}\nonumber\\
    =\,& \frac{\beta q e^{\beta a}e^{\beta b}+\beta e^{\beta b}(2e^{\beta a}+(q-2)e^{\beta b})+(q-2)\beta e^{\beta b}(e^{\beta a}-e^{\beta b})}{(e^{\beta a}+(q-1)e^{\beta b})^2}=\frac{2\beta q e^{\beta a} e^{\beta b}}{(e^{\beta a}+(q-1) e^{\beta b})^2}\nonumber\\
    =  \,  & \frac{2\beta q a b}{(a+(q-1)b)^2} = 2 q \beta  ab,
\end{align}
where we use \eqref{Eq2.3} in the last two equalities. Next, if $i\geq 2$, then $i,j\in [q]\backslash \{1\}$, and using \eqref{Eqq2}--\eqref{Eqq1}, we get
\begin{align}\label{Eq2.16}
      &\sum_{\ell=1}^q \bigg|\frac{\partial \Phi_{\ell}(\bm{\alpha}_0)}{\partial \alpha_i}-\frac{\partial \Phi_{\ell}(\bm{\alpha}_0)}{\partial \alpha_j}\bigg|=\bigg|\frac{\partial \Phi_i(\bm{\alpha}_0)}{\partial \alpha_i}-\frac{\partial \Phi_i(\bm{\alpha}_0)}{\partial \alpha_j}\bigg|+\bigg|\frac{\partial\Phi_j(\bm{\alpha}_0)}{\partial \alpha_i}-\frac{\partial \Phi_j(\bm{\alpha}_0)}{\partial \alpha_j}\bigg|\nonumber\\
      =\,& \frac{2|\beta e^{\beta b} (e^{\beta a}+(q-2)e^{\beta b})+\beta e^{2\beta b}|}{(e^{\beta a}+ (q-1)e^{\beta b})^2}
     =\frac{2\beta e^{\beta b}}{e^{\beta a}+(q-1)e^{\beta b}}=\frac{2\beta b}{a+(q-1)b}=2\beta b\leq 2q\beta ab,
\end{align}
where we use \eqref{Eq2.3} in the last two equalities and Lemma \ref{L2.1}(a) in the inequality in the second line.

Below we consider any $\bm{\alpha},\bm{\alpha}'\in\mathbb{R}^q$ such that $\sum_{\ell=1}^q \alpha_{\ell}=\sum_{\ell=1}^q \alpha_{\ell}'$. For any $\ell\in [q]$, let
\begin{equation}\label{Hexpressions}
    H_{\ell}(\bm{\alpha},\bm{\alpha}'):=\sum_{\ell'=1}^q (\alpha_{\ell'}-\alpha'_{\ell'})\frac{\partial \Phi_{\ell}(\bm{\alpha}_0)}{\partial \alpha_{\ell'}}. 
\end{equation}
We bound $\sum_{\ell=1}^q |H_{\ell}(\bm{\alpha},\bm{\alpha}')|$ as follows. Without loss of generality, we assume that $\bm{\alpha}\neq\bm{\alpha}'$. Let $I_{+}:=\{\ell\in [q]: \alpha_{\ell}>\alpha'_{\ell}\}$ and $I_{-}:=\{\ell\in [q]: \alpha_{\ell}<\alpha'_{\ell}\}$. As $\sum_{\ell\in I_{+}}(\alpha_{\ell}-\alpha'_{\ell})=\sum_{\ell\in I_{-}}(\alpha'_{\ell}-\alpha_{\ell})$, there exists a finite sequence $\{\bm{\alpha}_j\}_{j=0}^J$ (where $J\in\mathbb{N}^{*}$) from $\mathbb{R}^q$, such that $\bm{\alpha}_0=\bm{\alpha}$, $\bm{\alpha}_J=\bm{\alpha}'$, and for any $j\in [0,J-1]\cap\mathbb{N}$, there exists some $\ell_j,\ell_j'\in [q]$, such that $\ell_j\neq\ell_j'$, $(\bm{\alpha}_j)_{\ell_j}>(\bm{\alpha}_{j+1})_{\ell_j}$, $(\bm{\alpha}_j)_{\ell_j'}<(\bm{\alpha}_{j+1})_{\ell_j'}$, $(\bm{\alpha}_j)_{\ell_j}-(\bm{\alpha}_{j+1})_{\ell_j}=(\bm{\alpha}_{j+1})_{\ell_j'}-(\bm{\alpha}_j)_{\ell_j'}$, and $(\bm{\alpha}_j)_{\ell}=(\bm{\alpha}_{j+1})_{\ell}$ for any $\ell\in [q]\backslash \{\ell_j,\ell_j'\}$. Moreover, letting $s_j=(\bm{\alpha}_j)_{\ell_j}-(\bm{\alpha}_{j+1})_{\ell_j}$, we can select the sequence so that
\begin{equation}\label{Eq2.17}
    \sum_{j=0}^{J-1}s_j=\frac{1}{2}\sum_{\ell=1}^q |\bm{\alpha}_{\ell}-\bm{\alpha}'_{\ell}|. 
\end{equation}
For any $j\in [0,  J-1]\cap\mathbb{N}$, using \eqref{Hexpressions} we have 
\begin{equation*}
    \sum_{\ell=1}^q |H_{\ell}(\bm{\alpha}_j,\bm{\alpha}_{j+1})|= s_j \sum_{\ell=1}^q \bigg|\frac{\partial \Phi_{\ell}(\bm{\alpha}_0)}{\partial \alpha_{\ell_j}}-\frac{\partial \Phi_{\ell}(\bm{\alpha}_0)}{\partial \alpha_{\ell_j'}}\bigg|\leq 2q\beta ab s_j,
\end{equation*}
where the last inequality uses \eqref{Eq2.15} and \eqref{Eq2.16}. Hence by the above display and \eqref{Eq2.17}, we have 
\begin{equation*}\label{Eq2.19}
 \sum_{\ell=1}^q |H_{\ell}(\bm{\alpha},\bm{\alpha}')|\leq \sum_{j=0}^{J-1}  \sum_{\ell=1}^q |H_{\ell}(\bm{\alpha}_j,\bm{\alpha}_{j+1})|\leq 2q\beta a b\sum_{j=0}^{J-1} s_j=q\beta a b\sum_{\ell=1}^q |\bm{\alpha}_{\ell}-\bm{\alpha}'_{\ell}|.
\end{equation*}

We define $\phi_{\ell}(\theta)$ for any $\ell\in [q]$ and $\theta\in [0,1]$ as in (\ref{def_p}). For any $\ell\in [q]$,  
\begin{eqnarray*}\label{Eq2.20}
     \Phi_{\ell}(\bm{\alpha})-\Phi_{\ell}(\bm{\alpha}')&=&\phi_{\ell}(1)-\phi_{\ell}(0)=\int_{0}^{1} \phi_{\ell}'(\theta) d\theta\nonumber\\
    &=&\sum_{\ell'=1}^q (\alpha_{\ell'}-\alpha'_{\ell'})\int_{0}^{1} \frac{\partial \Phi_{\ell}(\theta \bm{\alpha}+(1-\theta)\bm{\alpha}')}{\partial \alpha_{\ell'}}d\theta.
\end{eqnarray*}
By Lemma \ref{L2.1}(a), $q\beta ab<1$. The conclusion of the lemma follows from the above two displays, on using the continuity of $\nabla \Phi$.  

\subsection{Proof of Lemma \ref{L2.4}}

Note that $\Phi_1(\bm{\alpha})-\Phi_2(\bm{\alpha})=\frac{e^{\beta \alpha_1}-e^{\beta \alpha_2}}{\sum_{k=1}^q e^{\beta \alpha_k}}\geq \frac{e^{\beta \alpha_1}-e^{\beta \alpha_2}}{e^{\beta \alpha_1}+(q-1)e^{\beta \alpha_2}}=g(\alpha_1-\alpha_2)$, where $g$ is defined as in Lemma~\ref{L2.1}(b). To complete the proof, note that for any $x\geq 0$,
\begin{equation*}
    g'(x)=\frac{q\beta e^{\beta x}}{(e^{\beta x}+q-1)^2}>0,
\end{equation*}
hence $g'(0)=\beta \slash q$ and $g$ is monotone increasing on $[0,\infty)$. Moreover, we have $g(0)=0$ (see Lemma~\ref{L2.1}(b)).
Let $\delta=\delta_{\beta,q,\rho}>0$ be such that $g'(x)>\rho \beta\slash q$ for $x\in [0,\delta]$, the existence of which follows from the continuity of $g'$. Then, if $\alpha_1-\alpha_2<\delta$, we have 
\begin{align*}
    g(\alpha_1-\alpha_2)=g(\alpha_1-\alpha_2)-g(0)\geq \rho\beta q^{-1}(\alpha_1-\alpha_2).
\end{align*}
On the other hand, if $\alpha_1-\alpha_2\geq\delta$, then
\begin{align*}
    g(\alpha_1-\alpha_2)\geq g(\delta)=:\kappa_{\rho},
\end{align*}
where $\kappa_{\rho}$ depends only on $q,\beta,\rho$. The conclusion of the lemma follows by combining the above two displays. 


\section{Specialized Auxiliary Gaussian algorithm for $q=2$}

\subsection{Standard Auxiliary Gaussian algorithm}

Setting $\mathbf W:=\mathbf Z_1-\mathbf Z_2$  and using \eqref{eq:main}, under $\mathbb P$ we have
\begin{align}\label{eq:main_ising}
(\mathbf W|\mathbf X=\mathbf x)\sim N(\mathbf y_1-\mathbf y_2, 2B^{-1}).
\end{align}
Also, using \eqref{eq:conditional}, we have
\begin{align}\label{eq:conditional_ising}
\notag\mathbb P(X_i=1|\mathbf Z=\mathbf z)=&\frac{\exp\Big(\sum_{j=1}^nB_{ij}z_{1j}\Big)}{\exp\Big(\sum_{j=1}^nB_{ij} z_{1j}\Big)+\exp\Big(\sum_{j=1}^nB_{ij} z_{2j}\Big)}\\
=&\frac{\exp\Big(\frac{1}{2}\sum_{j=1}^nB_{ij}w_j\Big)}{\exp\Big(\frac{1}{2}\sum_{j=1}^nB_{ij} w_j\Big)+\exp\Big(-\frac{1}{2}\sum_{j=1}^nB_{ij} w_j\Big)},
\end{align}
where $w_j=z_{1j}-z_{2j}$. A similar calculation gives
\begin{align*}
\mathbb P(X_i=2|\mathbf Z=\mathbf z)
 = \frac{\exp\Big(-\frac{1}{2}\sum_{j=1}^nB_{ij} w_j\Big)}{\exp\Big(\frac{1}{2}\sum_{j=1}^nB_{ij}w_j\Big)+\exp\Big(-\frac{1}{2}\sum_{j=1}^nB_{ij}w_j\Big)}.
\end{align*}
Thus, instead of using the $2n$ dimensional Gaussian ${\mathbf Z}$, one can use the $n$ dimensional Gaussian $W$, and iterate between \eqref{eq:main_ising} and \eqref{eq:conditional_ising}, This is the same exact algorithm, but we work with a lower dimensional representation of the auxiliary variable, which helps in faster computations. We note that this is the exact same algorithm as \citep{Martens:2010} in the Ising case.

\subsection{Low rank Auxiliary Gaussian algorithm}

Setting $\mathbf W:=\mathbf Z_1-\mathbf Z_2$  as before, using \eqref{eq:main2} under $\mathbb Q$ we have
\begin{align}\label{eq:main_ising_lr}
(\mathbf W_j|\mathbf X=\mathbf x)\sim N(\mathbf p_j'(\mathbf y_1-\mathbf y_2), 2/\mu_j),
\end{align}
with $(W_1,\ldots,W_k)$ mutually independent.
Also, using \eqref{eq:low_rank_conditional}, we have
\begin{align}\label{eq:conditional_ising_lr}
\notag\mathbb Q(X_i=1|\mathbf Z=\mathbf z)=&\frac{\exp\Big(\sum_{j=1}^k\mu_j z_{1j} p_{ji}\Big)}{\exp\Big(\sum_{j=1}^k\mu_j z_{1j} p_{ji}\Big)+\exp\Big(\sum_{j=1}^k\mu_j z_{2j} p_{ji}\Big)}\\
=&\frac{\exp\Big(\frac{1}{2}\sum_{j=1}^k\mu_j w_jp_{ji}\Big)}{\exp\Big(\frac{1}{2}\sum_{j=1}^k\mu_j w_j p_{ji}\Big)+\exp\Big(-\frac{1}{2}\sum_{j=1}^k\mu_j w_j p_{ji}\Big)},
\end{align}
where $w_j=z_{1j}-z_{2j}$. A similar calculation gives
\begin{align*}
\mathbb Q(X_i=2|\mathbf Z=\mathbf z)
=\frac{\exp\Big(-\frac{1}{2}\sum_{j=1}^k\mu_j w_jp_{ji}\Big)}{\exp\Big(\frac{1}{2}\sum_{j=1}^k\mu_j w_j p_{ji}\Big)+\exp\Big(-\frac{1}{2}\sum_{j=1}^k\mu_j w_j p_{ji}\Big)}.
\end{align*}
Thus, instead of using the $2k$ dimensional Gaussian ${\mathbf Z}$, one can use the $k$ dimensional Gaussian $W$, and iterate between \eqref{eq:main_ising_lr} and \eqref{eq:conditional_ising_lr}.

\bibliography{ref.bib}

\end{document}